\documentclass[11pt,a4paper]{llncs}
\usepackage{graphicx}
\usepackage{algpseudocode}
\usepackage{color}
\usepackage{amsmath}
\usepackage{amssymb}
\usepackage{amsfonts}
\usepackage{sidecap}
\usepackage{url}
\usepackage{multicol}
\usepackage{makecell}
\usepackage[left=0.885in,top=1in]{geometry}
\usepackage{xcolor}
\usepackage[floatrowsep=qquad,subfloatrowsep=qquad]{floatrow}
\usepackage[font={scriptsize}]{caption}
\usepackage{sidecap}

\def\SA{\mbox{\rm {\sf SA}}}

\def\LF{\mbox{\rm {\sf LF}}}

\def\rank{\textsf{rank}}

\def\BWT{\mbox{\rm {\sf BWT}}}

\usepackage{sidecap}
\usepackage{subfigure,url}
\usepackage{xspace}
\newcommand{\OCC}{\ensuremath{\mathrm{Occ}}}
\newcommand{\IL}{\ensuremath{\mathrm{IL}}}
\newcommand{\PP}{\ensuremath{\mathrm{EP}}}
\newcommand{\Prec}{\ensuremath{\mathrm{PR}}}
\newcommand{\SAT}{\ensuremath{\mathsf{SA}}}
\newcommand{\SAD}{\ensuremath{\mathsf{SA}_{\Dcal}}}
\newcommand{\BWTP}{\ensuremath{\mathsf{BWT}_{\Pcal}}}

\algnewcommand\KwTo{\textbf{to }} 
\algnewcommand\KwAnd{\textbf{and }}
\algnewcommand\KwWrite{\textbf{write }}
\algnewcommand\KwTofile{\textbf{to file }}
\long\def\ignore#1{}

\makeatletter
\newcommand{\printfnsymbol}[1]{%
  \textsuperscript{\@fnsymbol{#1}}%
}
\makeatother

\begin{document}

\newcommand{\main}{representative\xspace}
\newcommand{\word}{w}
\newcommand{\sep}{\#}
\newcommand{\Dcal}{{\mathcal D}}
\newcommand{\Pcal}{{\mathcal P}}

\title{Efficient Construction of a Complete Index for Pan-Genomics Read Alignment}
\author{Alan Kuhnle\inst{1}\thanks{Equal contribution}
\and Taher Mun\inst{2}\printfnsymbol{1} \and
Christina Boucher\inst{1} \and
Travis Gagie\inst{3}
\and Ben Langmead\inst{2} \and \\
 Giovanni Manzini\inst{4} 
}
\institute{
Department of Computer and Information Science and Engineering, \\
%College of Engineering, \\
University of Florida, Gainesville, FL, \\~\\
\and
Department of Computer Science, \\
John Hopkins University, Baltimore, MD, \\~\\
\and
School of Computer Science and Telecommunications, \\
CeBiB and Universidad Diego Portales, Santiago, Chile \\~\\
\and
Department of Science and Technological Innovation, \\
University of Eastern Piedmont, Alessandria, Italy \\~\\
}
\onecolumn
\maketitle

%\fancyfoot[R]{\scriptsize{Copyright \textcopyright\ 2019 by SIAM\\
%Unauthorized reproduction of this article is prohibited}}

% All of these methods build an index from a given database (e.g., one or more reference genomes or draft genomes) and then use the index to query for subsequences that match to the database.  And w
 
\begin{abstract}
While short read aligners, which predominantly use the FM-index, are able to easily index one or a few human genomes, they do not scale well to indexing databases containing thousands of genomes.  To understand why, it helps to examine the main components of the FM-index in more detail, which is a rank data structure over the Burrows-Wheeler Transform ($\BWT$) of the string that will allow us to find the interval in the string's suffix array ($\SA$) containing pointers to starting positions of occurrences of a given pattern; second, a sample of the $\SA$ that --- when used with the rank data structure --- allows us access the $\SA$.   The rank data structure can be kept small even for large genomic databases, by run-length compressing the $\BWT$, but until recently there was no means known to keep the $\SA$ sample small without greatly slowing down access to the $\SA$.  Now that Gagie et al. (SODA 2018) have defined an $\SA$ sample that takes about the same space as the run-length compressed $\BWT$  ---  we have the design for efficient FM-indexes of genomic databases but are faced with the problem of building them.  In 2018 we showed how to build the $\BWT$ of large genomic databases efficiently (WABI 2018) but the problem of building Gagie et al.'s $\SA$ sample efficiently was left open.   We compare our approach to state-of-the-art methods for constructing the $\SA$ sample, and demonstrate that it is the fastest and most space-efficient method on highly repetitive genomic databases.  Lastly, we apply our method for indexing partial and whole human genomes, and show that it improves over Bowtie with respect to both memory and time.  \\

{\bf Availability: } We note that the implementation of our methods can be found here: \url{https://github.com/alshai/r-index}.
 \end{abstract}

\newpage

\section{Introduction} \label{sec:introduction}

% short read alignment
 The FM-index, which is a compressed subsequence index based on Burrows Wheeler transform ($\BWT$), is the primary data structure majority of short read aligners ---  including Bowtie \cite{bowtie}, BWA \cite{bwa} and SOAP2 \cite{soap}.  These aligners build a FM-index based data structure of sequences from a given genomic database, and then use the index to perform queries that find approximate matches of sequences to the database. And while these methods can easily index one or a few human genomes, they do not scale well to indexing the databases of thousands of genomes.  This is problematic in analysis of the data produced by consortium projects, which routinely have several thousand genomes.  

In this paper, we address this need by introducing and implementing an algorithm for efficiently constructing the FM-index.  This will allow for the FM-index construction to scale to larger sets of genomes. To understand the challenge and solution behind our method, it helps to examine the two principal components of the FM-index: first a rank data structure over the $\BWT$ of the string that enables us to find the interval in the string's suffix array ($\SA$) containing pointers to starting positions of occurrences of a given pattern (and, thus, compute how many such occurrences there are); second, a sample of the $\SA$ that, when used with the rank data structure, allows us access the $\SA$ (so we can list those starting positions). 
Searching with an FM-index can be summarized as follows: starting with the empty suffix, for each proper suffix of the given pattern we use rank queries at the ends of the $\BWT$ interval containing the characters immediately preceding occurrences of that suffix in the string, to compute the interval containing the characters immediately preceding occurrences of the suffix of length 1 greater; when we have the interval containing the characters immediately preceding occurrences of the whole pattern, we use a $\SA$ sample to list the contexts of the corresponding interval in the $\SA$, which are the locations of those occurrences.
 
%The only previous attempt to apply this paradigm directly to the problem of indexing genomic databases, by M\"akinen et al.~\cite{Makinen2010} (see also~\cite{Sir12}), was hampered by the fact that the product of size of the $\SA$ sample and the time to access an element of the $\SA$ with it, grew at least linearly with the size of the database.
 Although it is possible to use a compressed implementation of the rank data structure that does not become much slower or larger even for thousands of genomes, the same cannot be said for the $\SA$ sample.  
The product of the size and the access time must be at least linear in the length of the string for the standard $\SA$ sample.  This implies that the FM-index will become much slower and/or much larger as the number of genomes in the databases grows significantly.  This bottleneck has forced researchers to consider variations of FM-indexes adapted for massive genomic datasets, such as Valenzuela et al.'s pan-genomic index \cite{Valenzuela2018} or Garrison et al.'s variation graphs \cite{Garrison}.  
Some of these proposals use elements of the FM-index, but all deviate in substantial ways from the description above.  Not only does this mean they lack the FM-index's long and successful track record, it also means they usually do not give us the BWT intervals for all the suffixes as we search (whose lengths are the suffixes' frequencies, and thus a tightening sequence of upper bounds on the whole pattern's frequency), nor even the final interval in the suffix array (which is an important input in other string processing tasks).

Recently, Gagie, Navarro and Prezza~\cite{Travis18}  proposed a different approach to $\SA$ sampling, that takes space proportional to that of the compressed rank data structure while still allowing reasonable access times.    While their result yields a potentially practical FM-index on massive databases, it does not directly lead to a solution since the problem of how to efficiently construct the $\BWT$ and $\SA$ sample remained open.  In a direction toward to fully realizing the theoretical result of Gagie et al.~\cite{Travis18}, 
Boucher et al. \cite{boucher2018} showed how to build the $\BWT$ of large genomic databases efficiently.
We refer to this construction as {\em prefix-free parsing}. It takes as input string $S$, and in one-pass generates a dictionary and a parse of $S$ with the property that the $\BWT$ can be constructed from dictionary and parse using workspace proportional to their total size and $O(|S|)$ time.  Yet, the resulting  index of Boucher et al. \cite{boucher2018}  has no $\SA$ sample, and therefore, only supports counting and not locating. This makes this index not directly applicable to many bioinformatic applications, such as sequence alignment.   %The problem of building Gagie et al.'s $\SA$ sample efficiently was left open.   In this paper, we solve this latter problem and thus, show that we can build the $\BWT$ and Gagie et al.'s $\SA$ sample together in roughly the same time and memory needed to construct the $\BWT$ alone.  

%In a direction toward to fully realizing the theoretical result of Gagie et al.~\cite{Travis18}, Boucher et al. \cite{boucher2018} devised a means to efficiently construct the {\em run-length compressed FM-index (RLFM-index)} for 1,000 copies of human chromosome 19. 
 %We refer to this construction as {\em prefix-free parsing}. It takes as input string $S$, and in one-pass generates a dictionary and a parse of the test with the property that the BWT of $S$ can be constructed from dictionary and parse using workspace proportional to their total size and $O(|S|)$ time.  % --- this, in turn, allows the $r$-index to be constructed for large datasets \cite{boucher2018}.  Hence, we giver a manner for counting the number of occurrences of a given pattern in the string using prefix-free parsing but are limited only to counting, i.e., we cannot tell where they are without the SA sample of the string.

\noindent{\bf Our contributions.}   In this paper, we present a solution for building the FM-index\footnote{With the $\SA$ sample of Gagie et al. \cite{Travis18}, this index is termed the $r$-index.} for very large datasets by showing that we can build the $\BWT$ and Gagie et al.'s $\SA$ sample together in roughly the same time and memory needed to construct the $\BWT$ alone.  We note that this algorithm is also based on  prefix-free parsing.  Thus, we begin by describing how to construct the $\BWT$ from the prefix-free parse, and then show that it can be modified to build the $\SA$ sample in addition to the $\BWT$ in roughly the same time and space.  We implement this approach, and refer to the resulting implementation as {\tt bigbwt}.  We compare it to state-of-the-art methods for constructing the $\SA$ sample, and demonstrate that {\tt bigbwt} currently the fastest and most space-efficient method for constructing the $\SA$ sample on large genomic databases.

Next, we demonstrate the applicability of our method to short read alignment.  In particular, we compare the memory and time needed by our method to build an index for collections of chromosome 19 with that of Bowtie.   Through these experiments, we show that Bowtie was unable to build indexes for our largest collections (500 or more) because it exhausted memory, whereas our method was able to build indexes up to 1,000 chromosome 19s (and likely beyond).   At 250 chromosome 19 sequences, the our method required only about 2\% of the time and 6\% the peak memory of Bowtie's.   Lastly,  we demonstrate that it is possible to index collections of  whole human genome assemblies with sub-linear scaling as the size of the collection grows.

%%% Local Variables:
%%% mode: latex
%%% TeX-master: "recomb"
%%% End:

\noindent{\bf Related work.} The development of methods for building and the FM-index on large datasets is closely related to the development short-read aligners for pan-genomics --- an area where there is growing interest  \cite{Schneeberger2009,Danek2014,Gagie2015}.   Here, we briefly describe some previous approaches to this problem and detail its connection to the work in this paper.  We note that majority of pan-genomic aligners requiring building the FM-index for a population of genomes and thus, can increase proficiency by using the methods described in this paper. 

GenomeMapper \cite{Schneeberger2009}, the method of Danek et al. \cite{Danek2014}, and GCSA  \cite{Siren2014} represent the genomes in a population as a graph, and then reduce the alignment problem to finding a path within the graph.  Hence, these methods require all possible paths to be identified, which is exponential in the worst case.  Some of these methods --- such as the GCSA --- use the FM-index to store and query the graph and could capitalize on our approach by building the index in the manner described here.  Another set of approaches \cite{Makinen2010,Ferrada2014,Gagie2015,Valenzuela2017}
consider the reference pan-genome as the concatenation of individual genomes
and exploits redundancy by using a compressed index. The hybrid index
\cite{Ferrada2014} operates on a Lempel-Ziv compression of the reference pan-genome.
An input parameter $M$ sets the maximum length of reads that can be aligned; the
parameter $M$ has a large impact on the final size of the index. For this reason,
the hybrid index is suitable for short-read alignment only, although there have
been recent heuristic modifications to allow longer alignments \cite{Ferrada2018}.
In contrast, the $r$-index, of which we provide an implementation in this work, has no such length
limitation.  The most recent implementation of the hybrid index is CHIC \cite{Valenzuela2018}.   Although CHIC has support for counting multiple occurrences of a pattern within a genomic database, 
it is an expensive operation, namely $O(\ell \log \log n )$, where $\ell$ is the number of
occurrences in the databases and $n$ is the length of the database. 
However, the $r$-index is capable of counting all occurrences of a pattern of length $m$ in $O(m)$ time up to polylog factors.
There are a number of other approaches building off the hybrid index or similar ideas \cite{Danek2014,Wandelt2013}; for an extended discussion,
we refer the reader to the survey of Gagie and Puglisi \cite{Gagie2015}.

Finally, a third set of approaches \cite{Huang2013a,Maciuca2016} attempts to encode variants within a single reference genome. BWBBLE by Huang et al. \cite{Huang2013a} follows this by supplementing
the alphabet to indicate if multiple variants occur at a single location. This approach does not support counting of the number of variants matching a specific alignment; also, it suffers
from memory blow-up when larger structural variations occur.
%Next, a kernel sequence is created from the LZ parsing that replaces repetitive phrases
%with their inital and final $M$ characters, where $M$ is a user-defined input parameter; this kernel sequence
%is then indexed with a standard read aligner\footnote

%%% Local Variables:
%%% mode: latex
%%% TeX-master: "recomb"
%%% End:

%% BWBBLE

%% GenomeMapper
%% https://genomebiology.biomedcentral.com/articles/10.1186/gb-2009-10-9-r98

%% MuGI: https://journals.plos.org/plosone/article?id=10.1371/journal.pone.0109384
\section{Background}

\subsection{$\BWT$ and FM indexes} \label{sec:bwt}
Consider a string $S$ of length $n$ from a totally ordered
alphabet $\Sigma$, such that the last character of $S$ is lexicographically less
than any other character in $S$.  Let $F$ be the list of $S$'s characters sorted lexicographically by the suffixes starting at those characters, and let $L$ be 
the list of $S$'s characters sorted lexicographically 
by the suffixes starting immediately after those characters. The list
$L$ is termed the
Burrows-Wheeler Transform \cite{Burrows18} of $S$ and denoted
$\BWT$.
 %(The names $F$ and $L$ are standard for these lists.)  
If \(S [i]\) is in position $p$ in $F$ then \(S [i - 1]\) is in position $p$ in $L$.  Moreover, if \(S [i] = S [j]\) then \(S [i]\) and \(S [j]\) have the same relative order in both lists; otherwise, their relative order in $F$ is the same as their lexicographic order.  This means that if \(S [i]\) is in position $p$ in $L$ then, assuming arrays are indexed from 0 and $\prec$ denotes lexicographic precedence, in $F$ it is in position
$j_i = |\{h\,:\,S [h] \prec S[i]\}| + |\{h\,:\,L [h] = S [i],\ h \leq p\}| - 1$. The mapping $i \mapsto j_i$ is termed the $\LF$ mapping. 
Finally, notice that the last character in $S$ always appears first in $L$.  
By repeated application of the $\LF$ mapping, we can invert the $\BWT$,
that is, recover $S$ from $L$.
Formally, the \emph{suffix array} $\SA$ of the string $S$ is an array such that entry $i$
is the starting position in $S$ of the $i$th largest suffix in lexicographical order. 
The above definition of the $\BWT$ is equivalent to the following:
\begin{equation} \label{eq:bwtFromSA} %%Please don't remove this, I refer to it later
  \BWT [i] = S[ (\SA[i] - 1) \mod n ].
\end{equation}

The $\BWT$ was introduced as an aid to data compression: it moves characters followed by similar contexts together and thus makes many strings encountered in practice locally homogeneous and easily compressible.  Ferragina and Manzini~\cite{FM05} showed how the $\BWT$ may be used for \emph{indexing} a string $S$:
given a pattern $P$ of length $m < n$, find the number and location of all occurrences of $P$ 
within $S$.
If we know the range \(\BWT (S) [i..j]\) occupied by characters immediately preceding occurrences of a pattern $Q$ in $S$, then we can compute the range \(\BWT (S) [i'..j']\) occupied by characters immediately preceding occurrences of \(c Q\) in $S$, for any character $c \in \Sigma$, since
\begin{eqnarray*}
i' & = & |\{h\,:\,S [h] \prec c\}| + |\{h\,:\,S [h] = c, h < i\}|\\
j' & = & |\{h\,:\,S [h] \prec c\}| + |\{h\,:\,S [h] = c, h \leq j\}| - 1\,.
\end{eqnarray*}
Notice \(j' - i' + 1\) is the number of occurrences of \(c Q\) in $S$. The essential components of an FM-index for $S$ are, first, an array storing \(|\{h\,:\,S [h] \prec c\}|\) for each character $c$ and, second, a rank data structure for $\BWT$ that quickly tells us how often any given character occurs up to any given position\footnote{Given a sequence (string) $S[1,n]$ over an alphabet $\Sigma = \{1,\ldots,\sigma\}$, a character $c \in \Sigma $, and an integer
$i$, $\rank_c(S,i)$ is the number of times that $c$ appears in $S[1,i]$.}.   To be able to locate the occurrences of patterns in $S$ (in addition to just counting them), the FM-index uses
a sampled\footnote{\emph{Sampled} means that only some fraction of entries of the suffix array are stored.} suffix array of $S$ and a bit vector indicating the positions in $\BWT$ of the characters preceding the sampled suffixes.

\subsection{Prefix-free parsing} \label{sec:pfpbwt}

Next, we give an overview of prefix-free parsing, which produces a dictionary $\mathcal D$ and a parse $\mathcal P$ by sliding a window of fixed width through the input string $S$.  We refer the reader to Boucher et al. \cite{boucher2018} for the formal proofs and Section~\ref{sec:bwtalgo} for the algorithmic details. A rolling hash function identifies when substrings are parsed into elements of a dictionary, which is a set of substrings of $S$. Intuitively, for a repetitive string, the same dictionary phrases will be encountered frequently.   

We now formally define the dictionary $\mathcal D$ and parse $\mathcal P$.  Given a string\footnote{For technical reasons, the string $S$ must terminate with $w$ copies of lexicographically least $\$$ symbol. } $S$ of length $n$, window size $w \in \mathbb{N}$ and modulus $p \in \mathbb{N}$, we construct the dictionary $\mathcal D$ of substrings of $S$ and the parse $\mathcal P$ as follows. We let $f$ be a hash function on strings of length $w$, and let $\mathcal T$ be the sequence of substrings $W = S[s,s+w - 1]$ such that $f(W) = 0 \mod p$ or $W = S[0, w - 1]$ or $W = S[n - w + 1, n]$, ordered by initial position in $S$; let $\mathcal T = \left( W_1 = S[s_1,s_1 + w - 1], \ldots, W_k = [s_k,s_k + w - 1] \right)$.%\footnote{GIO: I like this way of presenting the parsing that doesn't use the \$ symbols. However, if the reader goes to~\cite{boucher2018} for the proofs, or to gitlab for the code, she will find different definitions. Shall we put a warning for this?}
By construction the strings
$$
S[s_1, s_{2} + w - 1], S[s_2, s_{3} + w - 1], \ldots, S[s_{k-1}, s_{k} + w - 1]
$$ form a parsing of $S$ in which each pair of consecutive strings $S[s_i, s_{i+1} + w - 1]$ and $S[s_{i+1}, s_{i+2} + w - 1]$ overlaps by exactly $w$ characters.  We define $\mathcal D = \{ S[s_i, s_{i+1} + w - 1] : 1 \le i < k \}$; that is, $\mathcal D$ consists of the set of the unique substrings $s$ of $S$ such that $|s| > w$ and the first and last $w$ characters of $s$ form consecutive elements in $\mathcal T$. If $S$ has many repetitions we expect that $|\Dcal| \ll k$.  With a little abuse of notation we define the parsing $\Pcal$ as the sequence of lexicographic ranks of substrings in $\Dcal$: $\mathcal P = \left( \text{rank}_{\mathcal D}( S[s_i, s_{i + 1} + w - 1] ) \right)_{i = 1}^{k-1}$. The parse $\mathcal P$ indicates how $S$ may be reconstructed using elements of $\mathcal D$.  The dictionary $\mathcal D$ and parse $\mathcal P$ may be constructed in one pass over $S$ in $O\left(n + | \mathcal D | \log \left| \Dcal \right| \right)$ time if the hash function $f$ can be computed in constant time.

%% We don't actually need this?  
%Boucher et al. \cite{boucher2018} showed there is an order-preserving surjective mapping from the set of all suffixes of $S$ to the set $\mathcal U$ of all suffixes of length $> w$ of elements of $\mathcal D$---with ambiguities being resolved by consulting suffixes of the parse $\mathcal P$.  This insight, along with the knowledge that  the $\BWT$ can be defined in terms of $\SA$ of $S$ (Eq. \ref{eq:bwtFromSA} of Section \ref{sec:bwt}), yields the following theorem:
%\begin{theorem}[\cite{boucher2018}]
%\label{thm:main}
%We can compute the $\BWT$ of $S$ from $\mathcal D$ and $\mathcal P$ using workspace proportional to sum of the total length of $\mathcal P$ and the elements of $\mathcal D$ and $O(n)$ time  when we can work in internal memory.\qed
%\end{theorem}

% Then for any two substrings $W_1$, $W_2$ of $S$ of length $w$, such that
% $f(W_1) = f(W_2) = 0 \mod p$ and no substring
% Let the \emph{window} be a contiguous substring of $S$ of length $w$; initially,
% $W = S[0,\ldots,w-1]$. 
% Then, in one pass over the string $S$, $f( W )$ is computed for each window;
% whenever $f(W) \mod p = 0$, $W$ terminates one dictionary word and begins
% another. 

\subsection{$r$-index locating}

Policriti and Prezza~\cite{PP18} showed that if we have stored $\SA [k]$ for each value $k$ such that $\BWT [k]$ is the beginning or end of a run (i.e., a maximal non-empty unary substring) in $\BWT$, and we know both the range $\BWT [i..j]$ occupied by characters immediately preceding occurrences of a pattern $Q$ in $S$ and the starting position of one of those occurrences of $Q$, then when we compute the range \(\BWT [i'..j']\) occupied by characters immediately preceding occurrences of \(c Q\) in $S$, we can also compute the starting position of one of those occurrences of $c Q$.  Bannai et al \cite{BGI18} then showed that even if we have stored only $\SA [k]$ for each value $k$ such that $\BWT [k]$ is the beginning of a run, then as long as we know $\SA [i]$ we can compute $\SA [i']$.

Gagie, Navarro and Prezza~\cite{Travis18} showed that if we have stored in a predecessor data structure $\SA [k]$ for each value $k$ such that $\BWT [k]$ is the beginning of a run in $\BWT $, with $\phi^{- 1} (\SA [k]) = \SA [k + 1]$ stored as satellite data, then given $\SA [h]$ we can compute $\SA [h + 1]$ in $O (\log \log n)$ time as
$\SA [h + 1] = \phi^{-1} (\mathsf{pred} (\SA [h])) + \SA [h] - \mathsf{pred} (\SA [h])\,,$ where $\mathsf{pred} (\cdot)$ is a query to the predecessor data structure.  Combined with Bannai et al.'s result, this means that while finding the range $\BWT [i..j]$ occupied by characters immediately preceding occurrences of a pattern $Q$, we can also find $\SA [i]$ and then report $\SA [i + 1..j]$ in $O ((j - i) \log \log n)$-time, that is, $O (\log \log n)$-time per occurrence.

Gagie et al.\ gave the name {$r$-index} to the index resulting from combining a rank data structure over the run-length compressed $\BWT$ with their $\SA$ sample, and Bannai et al.\ used the same name for their index.  Since our index is an implementation of theirs, we keep this name; on the other hand, we do not apply it to indexes based on run-length compressed $\BWT$s that have standard $\SA$ samples or no $\SA$ samples at all.

% \blue{GIO: we need a description of the locate mechanism to explain why the we are interested in the SA samples at the beginning and end of a run.}

%%% Local Variables:
%%% mode: latex
%%% TeX-master: "recomb"
%%% End:

\section{Methods}\label{sec:methods}

%\blue{GIO. Notation issues
%\begin{itemize}
%\item we consistently use $\SA$ and $\BWT$ (or $\BWT(S)$) for SA and BWT of the input $S$, and $\BWTP$ or $\SAD$ for BWT and SA of others strings. Shall we say explicitly that we are using this notation. shall we always ise $\BWT$ instead of $\BWT(S)$ for uniformity
%\item I changed some $s_i$ back to $t_i$ in section~\ref{sec:bwtalgo}. I did that on purpose because $s_i$'s are used above as integers and in section~\ref{sec:saalgo} as different integers. Better not to use them here as strings. 
%\item Dictionary elements are now called ``phrases'' everywhere. If you like it the can be changed to ``words'' with no conflict.
%\end{itemize}}

Here, we describe our algorithm for building the $\SA$ or the sampled $\SA$ from the prefix free parse of a input string $S$, which is used to build the $r$-index.  We first review the algorithm from~\cite{boucher2018} for building the $\BWT$ of $S$ from the prefix free parse.  Next, we show how to modify this construction to compute the $\SA$ or the sampled $\SA$ along with the $\BWT$.

\subsection{Construction of $\BWT$ from prefix-free parse}\label{sec:bwtalgo}

We assume we are given a prefix-free parse of $S[1..n]$ with window size $w$ consisting of a dictionary $\Dcal$ and a parse $\Pcal$. We represent the dictionary as a string $\Dcal[1..\ell] = t_1\sep t_2 \sep \cdots t_{d-1}\sep t_d\sep$ where $t_i$'s are the dictionary phrases in lexicographic order and $\sep$ is a unique separator. We assume we have computed the $\SA$ of $\Dcal$, denoted by $\SAD[1..\ell]$ in the following, and the suffix array of $\Pcal$, denoted $\BWTP$, and the array $\OCC[1,d]$ such that $\OCC[i]$ stores the number of occurrences in the parse of the dictionary phrase~$t_i$. These preliminary computations take $O(|\Dcal|+|\Pcal|)$ time. 

By the properties of the prefix-free parsing, each suffix of $S$ is prefixed by {\em exactly one} suffix $\alpha$ of a dictionary phrase $t_j$ with $|\alpha|>w$. We call $\alpha$ the {\em \main prefix} of the suffix $S[i..n]$. From the uniqueness of the \main prefix we can partition $S$'s suffix array $\SAT[1..n]$ into $k$ ranges
$$
[b_1,e_1],\quad [b_2,\ell_2],\quad [b_3, \ell_3],\quad \ldots,\quad [b_k,\ell_k]
$$
with $b_1=1$, $b_i=e_i+1$ for $i=2,\ldots,k$, and $e_k=n$, such that for $i=1,\ldots,k$ all suffixes 
$$
S[\SAT[b_i]..n],\quad S[\SAT[b_i+1]..n],\quad\ldots,\quad S[\SAT[e_i]..n]
$$
have the same \main\ prefix $\alpha_i$. By construction $\alpha_1 \prec \alpha_2 \prec \cdots \prec \alpha_k$. 

By construction, any suffix $\Dcal[i..\ell]$ of the dictionary $\Dcal$ is also prefixed by the suffix of a dictionary phrase. For $j=1,\ldots,\ell$, let $\beta_j$ denote the longest prefix of $\Dcal[\SAD[j]..\ell]$ which is the suffix of a phrase (i.e. $\Dcal[\SAD[j]+|\beta_j|]=\sep$). By construction the strings $\beta_j$'s are lexicographically sorted $\beta_1 \prec \beta_2 \prec \cdots \prec \beta_\ell$. Clearly, if we compute $\beta_1,\ldots,\beta_\ell$ and discard those such that $|\beta_j| \leq w$, the remaining $\beta_j$'s will coincide with the \main\ prefixes $\alpha_i$'s. Since both $\beta_j$'s and $\alpha_i$'s are lexicographically sorted, this procedure will generate the \main\ prefixes in the order $\alpha_1, \alpha_2, \ldots, \alpha_k$. We note that more than one $\beta_j$ can be equal to some $\alpha_i$ since different dictionary phrases can have the same suffix.

We scan $\SAD[1..\ell]$, compute $\beta_1, \ldots \beta_\ell$ and use these strings to find the \main\ prefixes. As soon as we generate an $\alpha_i$ we compute and output the portion $\BWT[b_i,e_i]$ corresponding to the range $[b_i,e_i]$ associated to $\alpha_i$. To implement the above strategy, assume there are exactly $k$ entries in $\SAD[1..\ell]$ prefixed by $\alpha_i$. This means that there are $k$ distinct dictionary phrases $t_{i_1}, t_{i_2}, \ldots, t_{i_k}$ that end with $\alpha_i$. Hence, the range $[b_i,e_i]$ contains 
$
z_i = e_i - b_i+1 = \sum_{h=1}^k \OCC[i_h]
$ 
elements. To compute $\BWT[b_i,e_i]$ we need to: 1) find the symbol immediately preceding each occurrence of $\alpha_i$ in $S$, and 2) find the lexicographic ordering of $S$'s suffixes prefixed by $\alpha_i$. We consider the latter problem first.  \vspace{-2mm}

\paragraph{Computing the lexicographic ordering of suffixes.}  For $j=1,\ldots,z_i$ consider the $j$-th occurrence of $\alpha_i$ in $S$ and let $i_j$ denote the position in the parsing of $S$ of the phrase ending with the $j$-th occurrence of $\alpha_i$. In other words, $\Pcal[i_j]$ is a dictionary phrase ending with $\alpha_i$ and $i_1 < i_2 < \cdots < i_{z_i}$. By the properties of $\BWTP$ the lexicographic ordering of $S$'s suffixes prefixed by $\alpha_i$ coincides with the ordering of the symbols $\Pcal[i_j]$ in $\BWTP$. In other words, $\Pcal[i_j]$ precedes $\Pcal[i_h]$ in $\BWTP$ if and only if $S$'s suffix prefixed by the $j$-th occurrence of $\alpha_i$ is lexicographically smaller than $S$'s suffix prefixed by the $h$-th occurrence of $\alpha_i$. 

We could determine the desired lexicographic ordering by scanning $\BWTP$ and noticing which entries coincide with one of the dictionary phrases $t_{i_1},\ldots, t_{i_k}$ that end with $\alpha_i$ but this would clearly be inefficient. Instead, for each dictionary phrase $t_i$ we maintain an array $\IL_i$ of length $\OCC[i]$ containing the indexes $j$ such that $\BWTP[j]=i$. These sorts of ``inverted lists'' are computed at the beginning of the algorithm and replace the $\BWTP$ which can be discarded. \vspace{-2mm}

\paragraph{Finding the symbol preceding $\alpha_i$.} Given a \main prefix $\alpha_i$ from $\SAD$ we retrieve the indexes $i_1, \ldots, i_k$ of the dictionary phrases $t_{i_1},\ldots, t_{i_k}$ that end with $\alpha_i$. Then, we retrieve the inverted lists $\IL_{i_1}, \ldots \IL_{i_k}$ and we merge them obtaining the list of the $z_i$ positions $y_1 < y_2 < \cdots < y_{z_i}$ such that $\BWT_P[y_j]$ is a dictionary phrase ending with $\alpha_i$. Such list implicitly provides the lexicographic order of $S$'s suffixes starting with $\alpha_i$. 

To compute the $\BWT$ we need to retrieve the symbols preceding such occurrences of $\alpha_i$. If $\alpha_i$ {\em is not} a dictionary phrase, then $\alpha_i$ is a proper suffix of the phrases $t_{i_1},\ldots, t_{i_k}$ and the symbols preceding $\alpha_i$ in $S$ are those preceding $\alpha_i$ in  $t_{i_1},\ldots, t_{i_k}$ that we can retrieve from $\Dcal[1..\ell]$ and $\SAD[1..\ell]$. If $\alpha_i$ {\em coincides} with a dictionary phrase $t_j$, then it cannot be a suffix of another phrase. Hence, the symbols preceding $\alpha_i$ in $S$ are those preceding $t_{j}$ in $S$ that we store at the beginning of the algorithm in an auxiliary array $\Prec_{j}$ along with the inverted list $\IL_{j}$.

\subsection{Construction of $\SA$ and $\SA$ sample along with the $\BWT$}\label{sec:saalgo}

We now show how to modify the above algorithm so that, along with $\BWT$, it computes the full $\SA$ of $S$ or the sampled $\SA$ consisting of the values $\SAT [s_1],\ldots,\SAT[s_r]$ and $\SAT[e_1],\ldots, \SAT [e_r]$, where $r$ is the number of maximal non-empty runs in $\BWT$ and $s_i$ and $e_i$ are the starting and ending positions in $\BWT$ of the $i$-th such run, respectively. Note that 
if we compute the sampled $\SA$ the actual output will consist of $r$ start-run pairs $\langle s_i, \SAT[s_i]\rangle$ and $r$ end-run pairs $\langle e_i, \SAT[e_i]\rangle$ since the $\SAT$ values alone are not enough for the construction of the $r$-index. 

We solve both problems using the following strategy. Simultaneously to each entry $\BWT[j]$, we compute the corresponding entry $\SAT[j]$.  Then, if we need the sampled $\SA$, we compare $\BWT[j-1]$ and $\BWT[j]$ and if they differ, we output the pair $\langle j-1, \SAT[j-1]\rangle$ among the end-runs and the pair $\langle j, \SAT[j]\rangle$ among the start-runs. To compute the $\SA$ entries, we only need $d$ additional arrays $\PP_1, \ldots \PP_d$ (one for each dictionary phrase), where $|\PP_i| = |\IL_i| = \OCC[i]$, and $\PP_i[j]$ contains the ending position in $S$ of the dictionary phrase which is in position $\IL_i[j]$ of $\BWTP$. 

Recall that in the above algorithm for each occurrence of a \main prefix $\alpha_i$, we compute the indexes $i_1,\ldots, i_k$ of the dictionary phrases $t_{i_1}, \ldots, t_{i_k}$ that end with $\alpha_i$. Then, we use the lists $\IL_{i_1}, \ldots, \IL_{i_k}$ to retrieve the positions of all the occurrences of $t_{i_1},\ldots, t_{i_k}$ in $\BWTP$, thus establishing the relative lexicographic order of the occurrences of the dictionary phrases ending with $\alpha_i$. To compute the corresponding $\SAT$ entries, we need the starting position in $S$ of each occurrence of $\alpha_i$. Since the ending position in $S$ of the phrase with relative lexicographic rank $\IL_{i_h}[j]$ is $\PP_{i_h}[j]$, the corresponding $\SA$ entry is $\PP_{i_h}[j]-|\alpha_i|+1$. Hence, along with each $\BWT$ entry we obtain the corresponding $\SA$ entry which is saved to the output file if the full $\SA$ is needed, or further processed as described above if we need the sampled $\SA$. 

\section{Time and memory usage for $\SA$ and $\SA$ sample construction} \label{sec:experiments}
We compare the running time and memory usage of {\tt bigbwt}\footnote{Our implementation of the algorithm in Section 3, available here: \url{https://gitlab.com/manzai/Big-BWT}.} with the following methods, which represent the current state-of-the-art. 
\begin{itemize}
  \item[{\tt bwt2sa}] Once the $\BWT$ has been computed, the $\SA$ or $\SA$ sample may
    be computed by applying the $\LF$ mapping to invert the $\BWT$ and the application of
    Eq. \ref{eq:bwtFromSA}. Therefore, as a baseline,
    we use {\tt bigbwt} to construct the $\BWT$ only, as in Boucher \emph{et al.} \cite{boucher2018};
    next, we load the $\BWT$ as a Huffman-compressed string with access, rank, and select support
    to compute the $\LF$ mapping. We step backwards through the $\BWT$ and compute
    %and we step backwards by computing the LF mapping. By exploiting the relationship
    %between $\BWT$ and $\SA$, we compute the $\SA$ or $\SA$ sample from this backwards stepping, which
    %generates 
    the entries of the $\SA$ in non-consecutive order. Finally, these entries
    are sorted in external memory to produce the $\SA$ or $\SA$ sample. This
    method may be parallelized when the input consists of multiple strings by stepping
    backwards from the end of each string in parallel.
  \item[{\tt pSAscan}] A second baseline is to compute the $\SA$ directly from the input;
    for this computation, we use the external-memory algorithm {\tt pSAscan}
    \cite{Karkkainen2015}, with available memory set to the memory required by {\tt bigbwt} on
    the specific input; with the ratio of memory to input size obtained from {\tt bigbwt}, 
    {\tt pSAscan} is the current state-of-the-art method to compute the $\SA$. Once 
    {\tt pSAscan} has computed the full $\SA$, the $\SA$ sample
    may be constructed by loading the input text $T$ into memory, streaming the
    $\SA$ from the disk, and the application of Eq. \ref{eq:bwtFromSA} to
    detect run boundaries.
    We denote this method of computing the
    $\SA$ sample by {\tt pSAscan+}.
  %\item[{\tt pSAscan+}]
\end{itemize}

We compared the performance of all the methods on two datasets: (1) Salmonella genomes obtained from GenomeTrakr \cite{STBASBM17}; and (2) chromosome 19
haplotypes derived from the 1000 Genomes Project phase 3 data \cite{1kg}.   The Salmonella strains were downloaded from NCBI (NCBI BioProject PRJNA183844) and preprocessed by assembling each individual sample with IDBA-UD \cite{Peng2012} and counting  $k$-mers ($k$=32) using KMC \cite{Deorowicz2015}.   We modified IDBA by setting  kMaxShortSequence
to $1024$ per public advice from the author to accommodate the longer paired end reads that modern sequencers produce.  We sorted the full set of samples by the size of their $k$-mer counts and selected 1,000 samples about the median.  This avoids exceptionally short assemblies, which may be due to low read coverage, and exceptionally long assemblies which may be due to contamination.  

Next, we downloaded and preprocessed a collection of chromosome 19 haplotypes from 1000 Genomes Project.  Chromosome 19 is 58 million base pairs in length and makes up around 1.9\% of the
total human genome sequence.  Each sequence was derived by using the
\texttt{bcftools consensus} tool to combine the haplotype-specific (maternal or
paternal) variant calls for an individual in the 1KG project with the chr19 sequence in
the GRCH37 human reference, producing a FASTA record per sequence.  All DNA characters besides A, C, G, T and N were removed from the sequences before
construction.

\begin{figure}[b] \centering
  \subfigure[Salmonella, 1 thread]{ \label{fig:SAtime1}
    \includegraphics[width=0.22\textwidth,height=0.13\textheight]{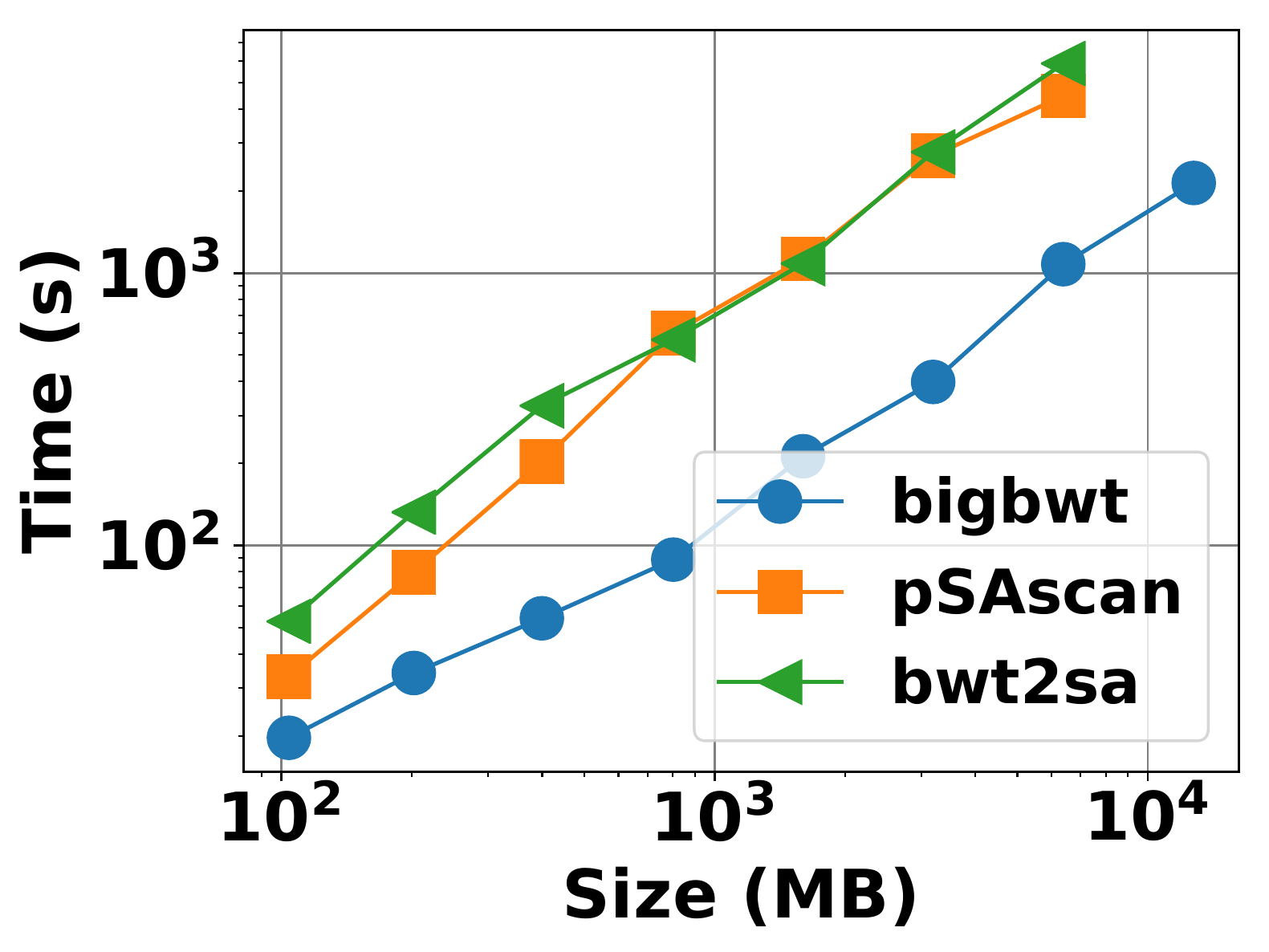} }
  %\subfigure[16 threads]{
  %\includegraphics[width=0.22\textwidth,height=0.13\textheight]{./salmonella-wall-time-16.pdf}
  % }
  \subfigure[chr19, 1 thread]{
    \includegraphics[width=0.22\textwidth,height=0.13\textheight]{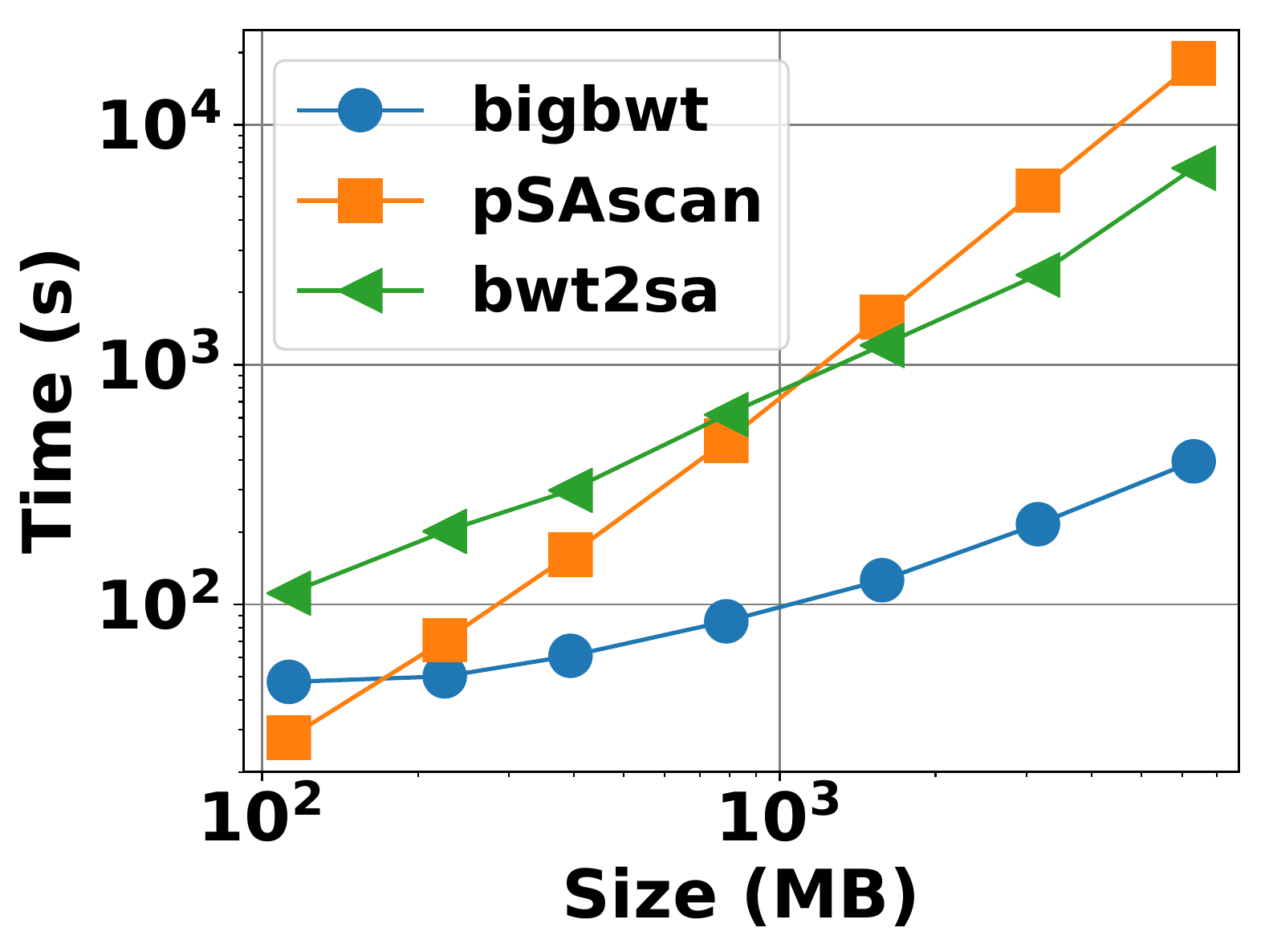}
  }
  \subfigure[chr19, 16 threads]{ \label{fig:SAtime3}
    \includegraphics[width=0.22\textwidth,height=0.13\textheight]{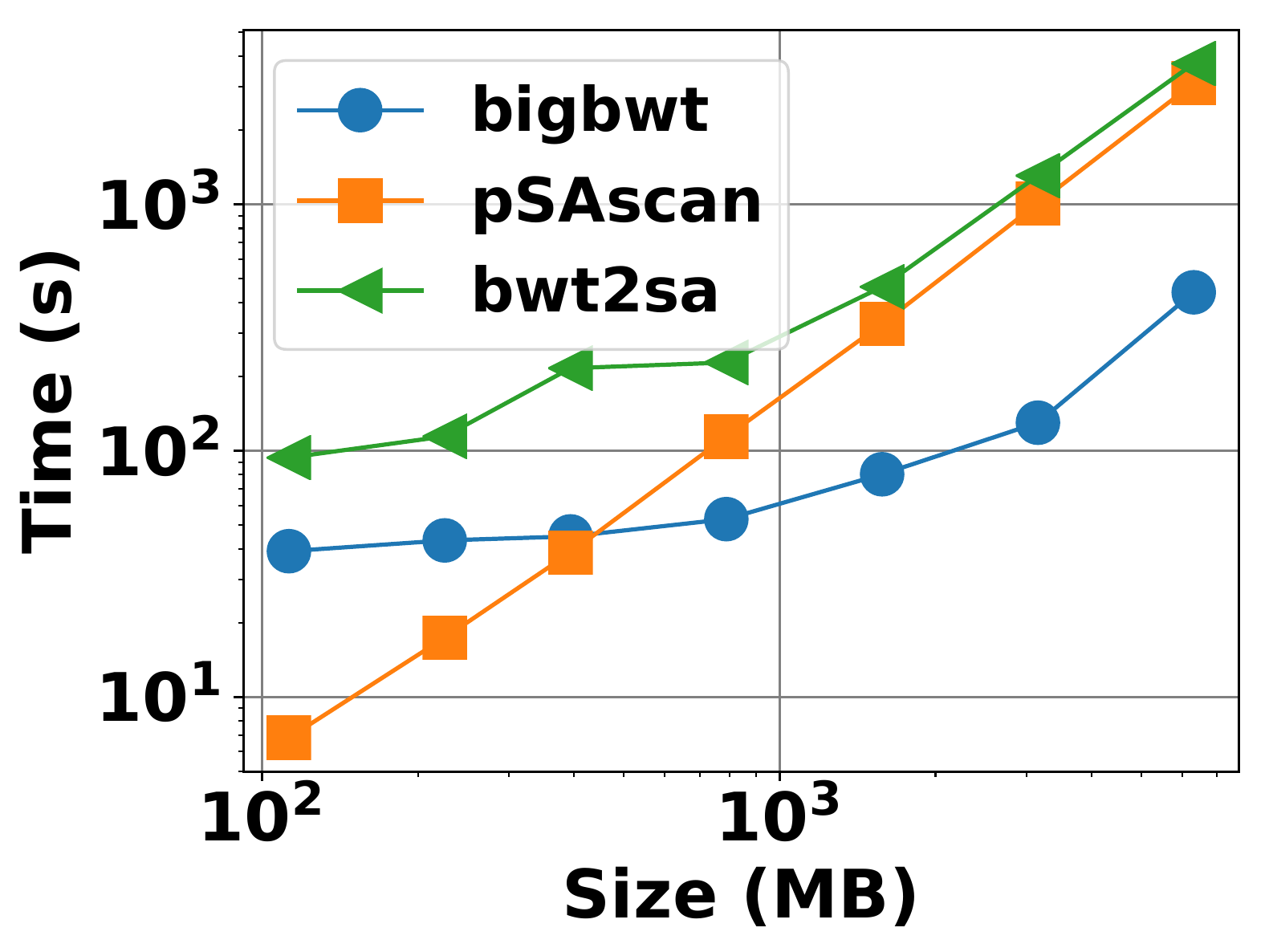}
  }  
  \subfigure[Peak memory usage]{
    \includegraphics[width=0.22\textwidth,height=0.13\textheight]{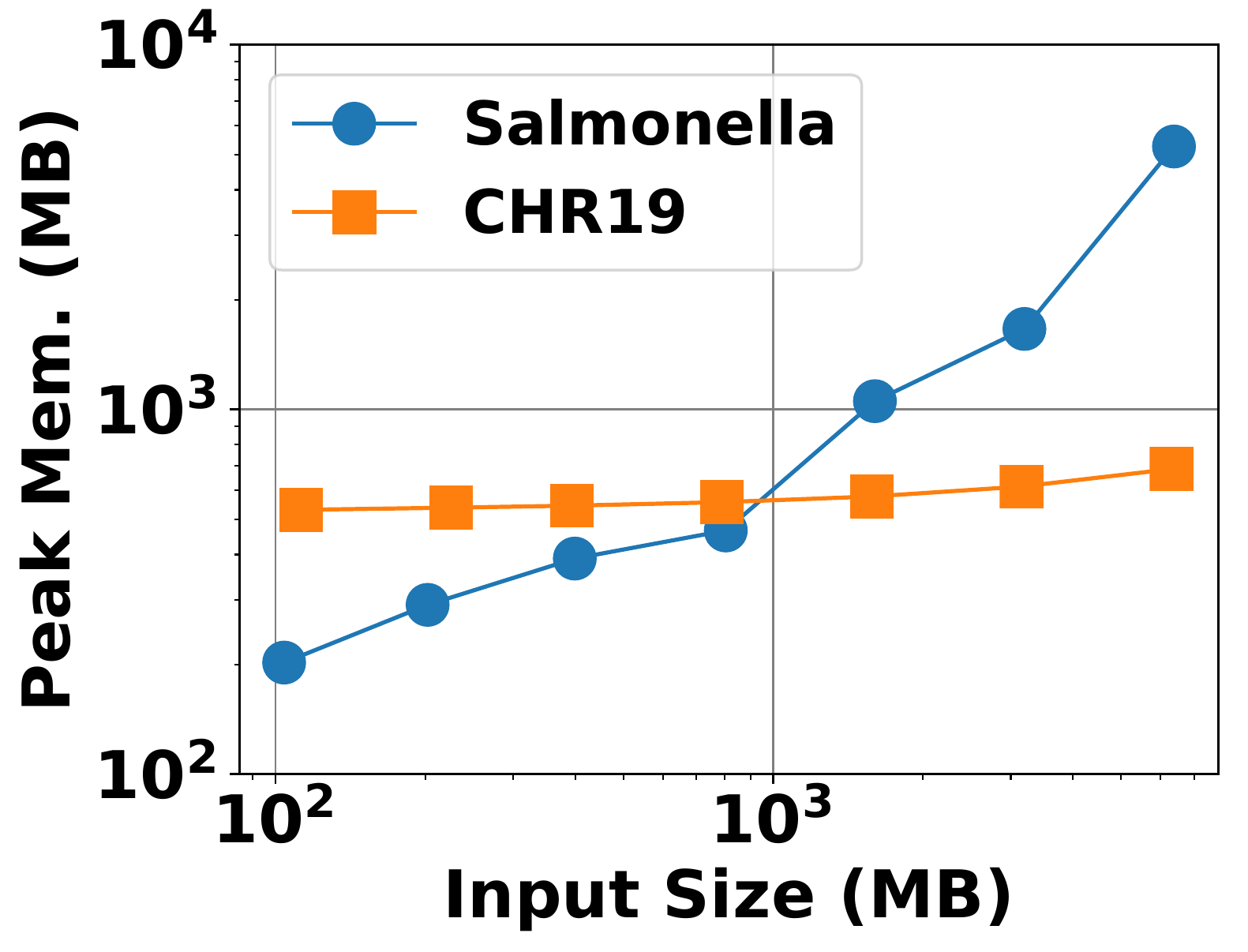} \label{fig:SAmem}
  }
  \caption{Runtime and peak memory usage for construction of full $\SA$.} \label{fig:fullSA}
\vspace{-3mm}
\end{figure}

\begin{figure}[h!]  \centering
  \subfigure[Salmonella, 1 thread]{
    \includegraphics[width=0.22\textwidth,height=0.13\textheight]{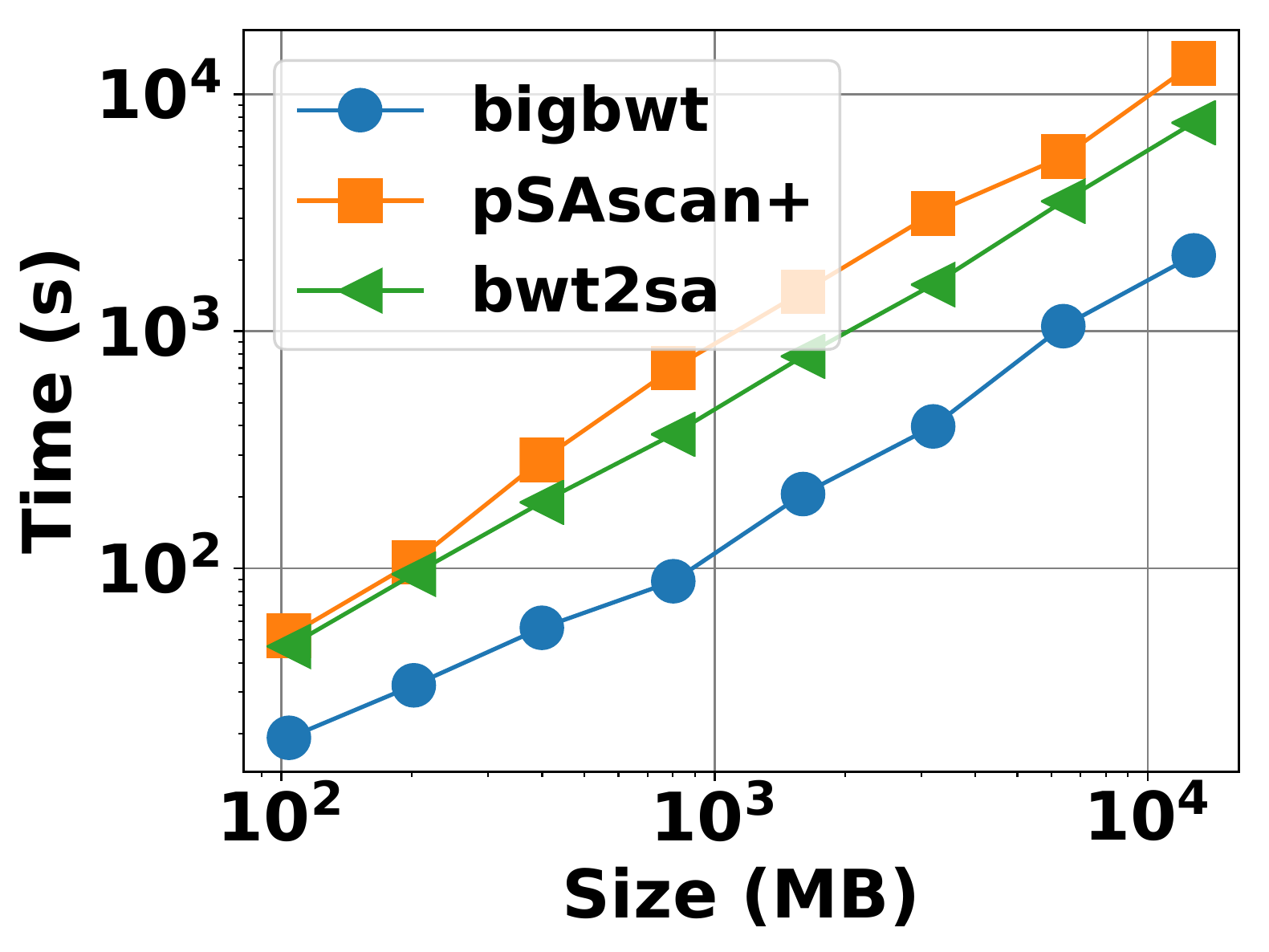} }
  %\subfigure[16 threads]{
  %\includegraphics[width=0.22\textwidth,height=0.13\textheight]{./salmonella-wall-time-16.pdf}
  % }
  \subfigure[chr19, 1 thread]{
    \includegraphics[width=0.22\textwidth,height=0.13\textheight]{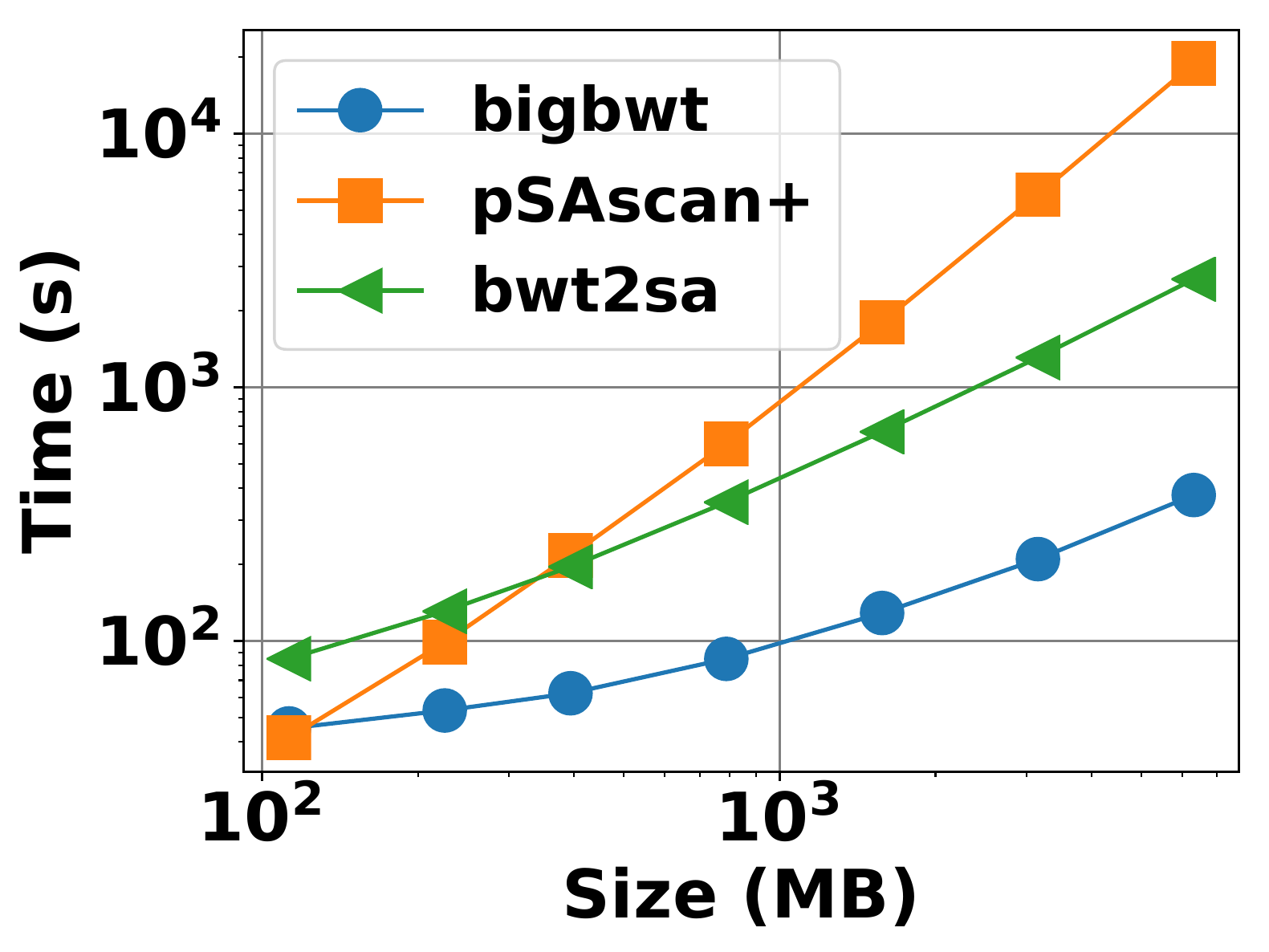}
  }
  \subfigure[chr19, 16 threads]{
    \includegraphics[width=0.22\textwidth,height=0.13\textheight]{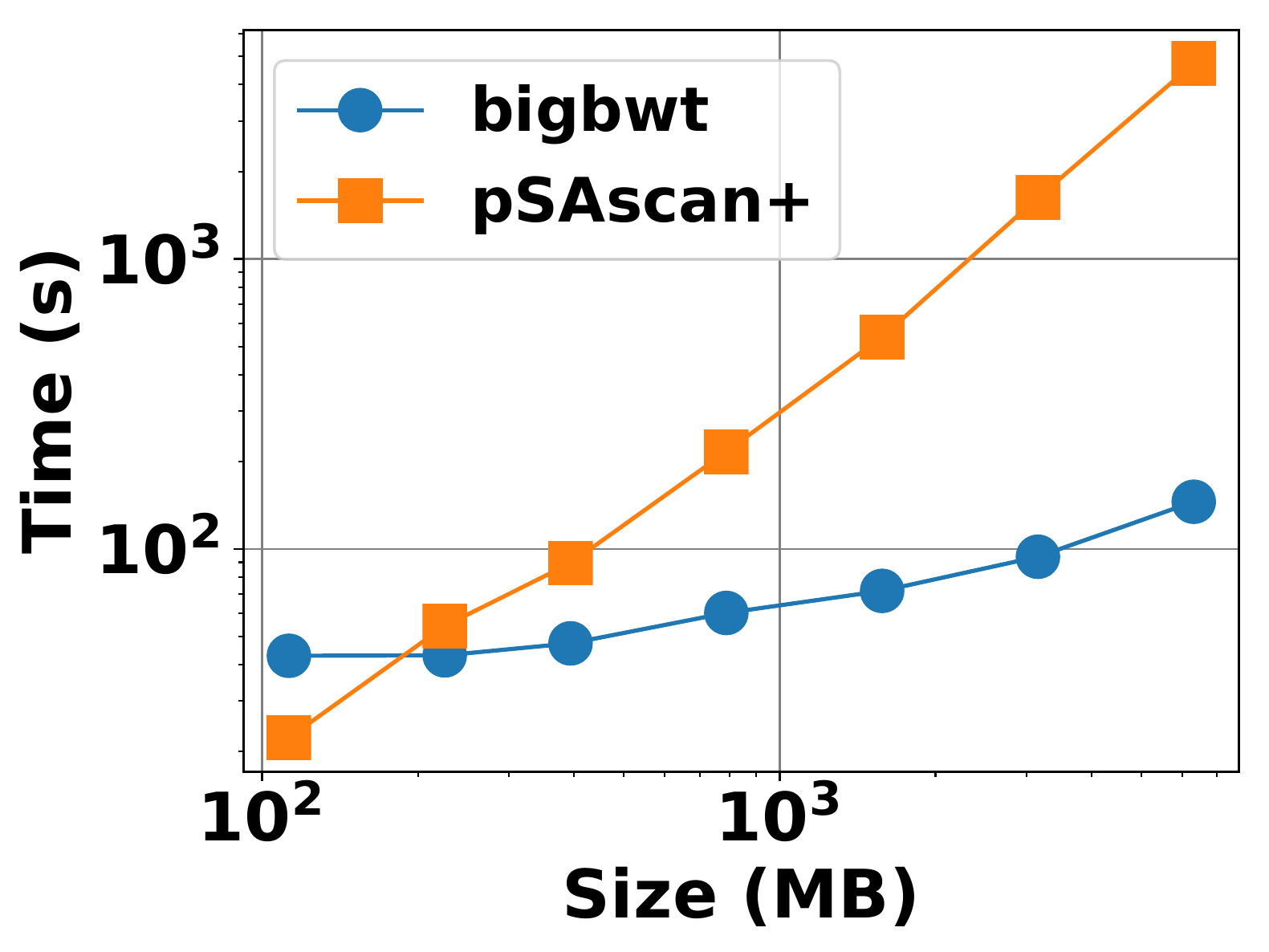}
  }  
  \subfigure[Peak memory usage]{
    \includegraphics[width=0.22\textwidth,height=0.13\textheight]{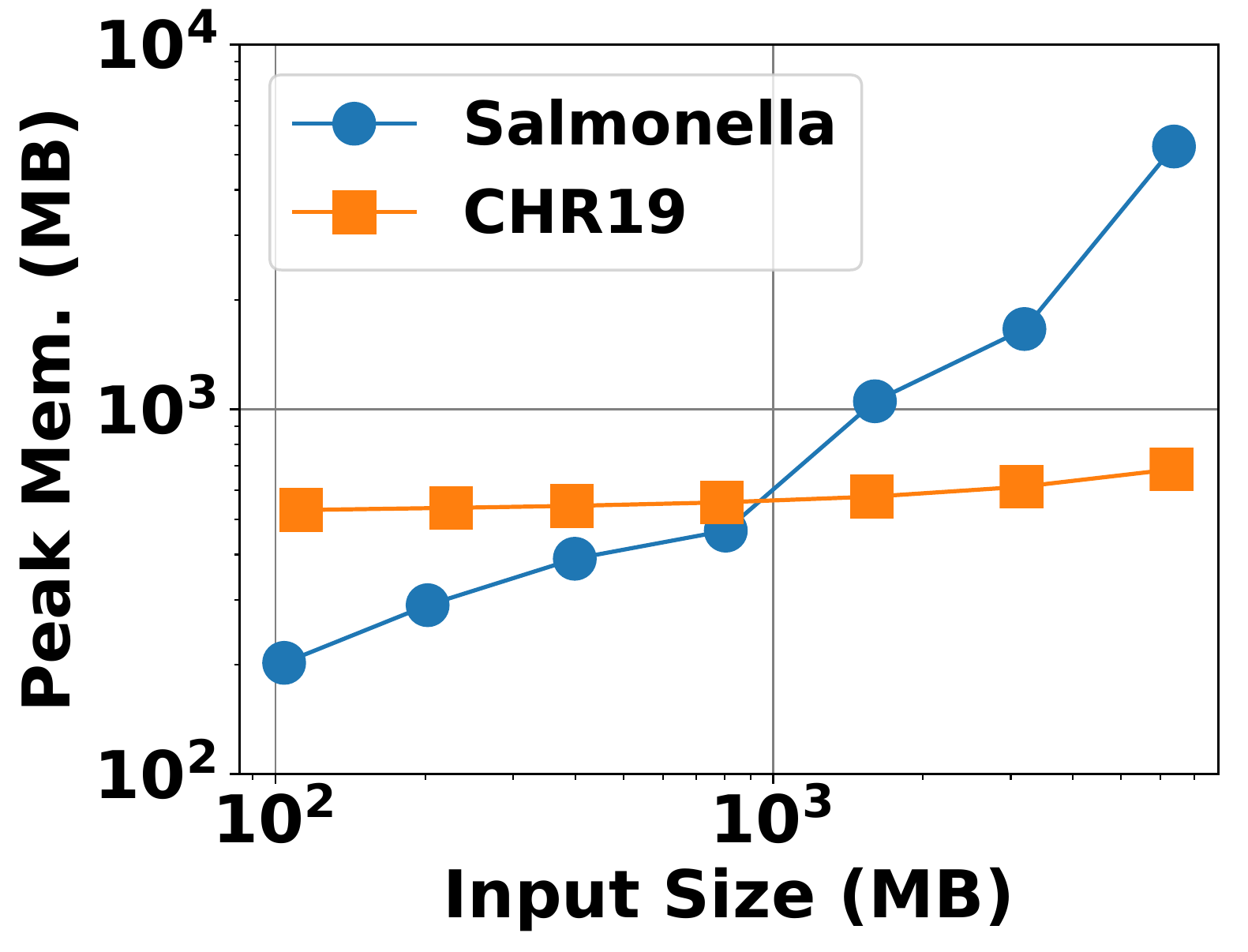} 
  }
  \caption{Runtime and peak memory usage for construction of $\SA$ sample.} \label{fig:SAsample}
\end{figure}

%\subsection{Competing Methods}

%\subsection{Results on Construction of SA (Sample)}
We performed all experiments in this section on a machine with Intel(R) Xeon(R) CPU E5-2680 v2 @ 2.80GHz and 324 GB RAM.  We measured running time and peak memory footprint using \texttt{/usr/bin/time -v}, with peak memory footprint captured by the \texttt{Maximum resident
set size (kbytes)} field and running time by the \texttt{User Time}
and \texttt{System Time} field.

We witnessed that the running time of each method to construct the full $\SA$ is shown in Figs. \ref{fig:SAtime1} -- \ref{fig:SAtime3}. On
both the Salmonella and chr19 datasets, {\tt bigbwt} ran the fastest, often by more
than an order of magnitude. In Fig. \ref{fig:SAmem}, we show the peak memory usage
of {\tt bigbwt} as a function of input size. Empirically, the peak memory usage 
was sublinear in input size, especially on the chr19 data, which
exhibited a high degree of repetition. Despite the higher diversity of the Salmonella genomes,
{\tt bigbwt} remained space-efficient and the fastest method for construction of 
the full $\SA$. Furthermore, we found qualitatively similar results for construction of the $\SA$ sample,
shown in Fig. \ref{fig:SAsample}. Similar to the results on full $\SA$ construction,
{\tt bigbwt} outperformed both baseline methods and exhibited 
sublinear memory scaling on both types of databases.  

%\paragraph{Results on construction of SA sample}
%%% Local Variables:
%%% mode: latex
%%% TeX-master: "recomb"
%%% End:

%        File: recomb.tex
%     Created: Thu Oct 25 04:00 PM 2018 E
% Last Change: Thu Oct 25 04:00 PM 2018 E
%

\section{Application to many human genome sequences} \label{sec:bowtie}

We studied how the $r$-index scales to repetitive texts
consisting of many similar genomic sequences.
Since an ultimate goal is to improve read alignment,
we benchmark against Bowtie (version 1.2.2) \cite{bowtie} .
%  FM index-based aligner as a benchmark.
%Popular FM-index aligners like Bowtie 2
%\cite{langmead_fast_2012} and BWA 
%use the locate query in an initial seed-finding step.
%This is followed by a seed extension step that uses dynamic programming to
%produce longer alignments that may include gaps and mismatches.
%Since our $r$-index implementation
%supports only exact matching, 
We ran Bowtie with the \texttt{-v 0} and \texttt{--norc} options;
\texttt{-v 0} disables approximate matching, while \texttt{--norc} causes 
Bowtie (like $r$-index) to perform the locate query with respect to the query sequence only
and not its reverse complement.
%Together, these options cause
%Bowtie to perform an comparable set of
%index probes and locate queries to those performed by $r$-index.
%.  We therefore compare the index construction and
%locate steps of our method against Bowtie 1 (version 1.2.2), which, just like
%RLFM, performs alignment through by backstepping through the BWT.

\subsection{Indexing chromosome 19s}

We performed our experiments on collections of one or more versions of chromosome 19.  
These versions were obtained from 1000 Genomes Project haplotypes in the manner described in the previous section.
We used 10
collections of chromosome 19 haplotypes, containing 1, 2, 10, 30, 50, 100, 250,
500, and 1000 sequences, respectively.
Each collection is a superset of the previous.
Again, all DNA characters besides A, C, G, T and N were removed from the sequences before
construction. 
All experiments in this section were ran on a Intel(R) Xeon(R) CPU E5-2680 v3 @ 2.50GHz
machine with 512GB memory. 
We measured running time and peak memory footprint as described in the previous section.  %using \texttt{/usr/bin/time -v}, with peak memory footprint captured by the \texttt{Maximum resident
%set size (kbytes)} field and running time by the \texttt{User Time}
%and \texttt{System Time} field.
%We also
%compare RLFM and Bowtie based on the disk footprint of their indexes.

% $r$-index uses prefix-free parsing to construct the
%$\BWT$ and the $\BWT$ run-ends and run-starts were computed during this step. 

\begin{figure}[t]
    \centering
  \subfigure[]{
     \includegraphics[width=0.23\textwidth]{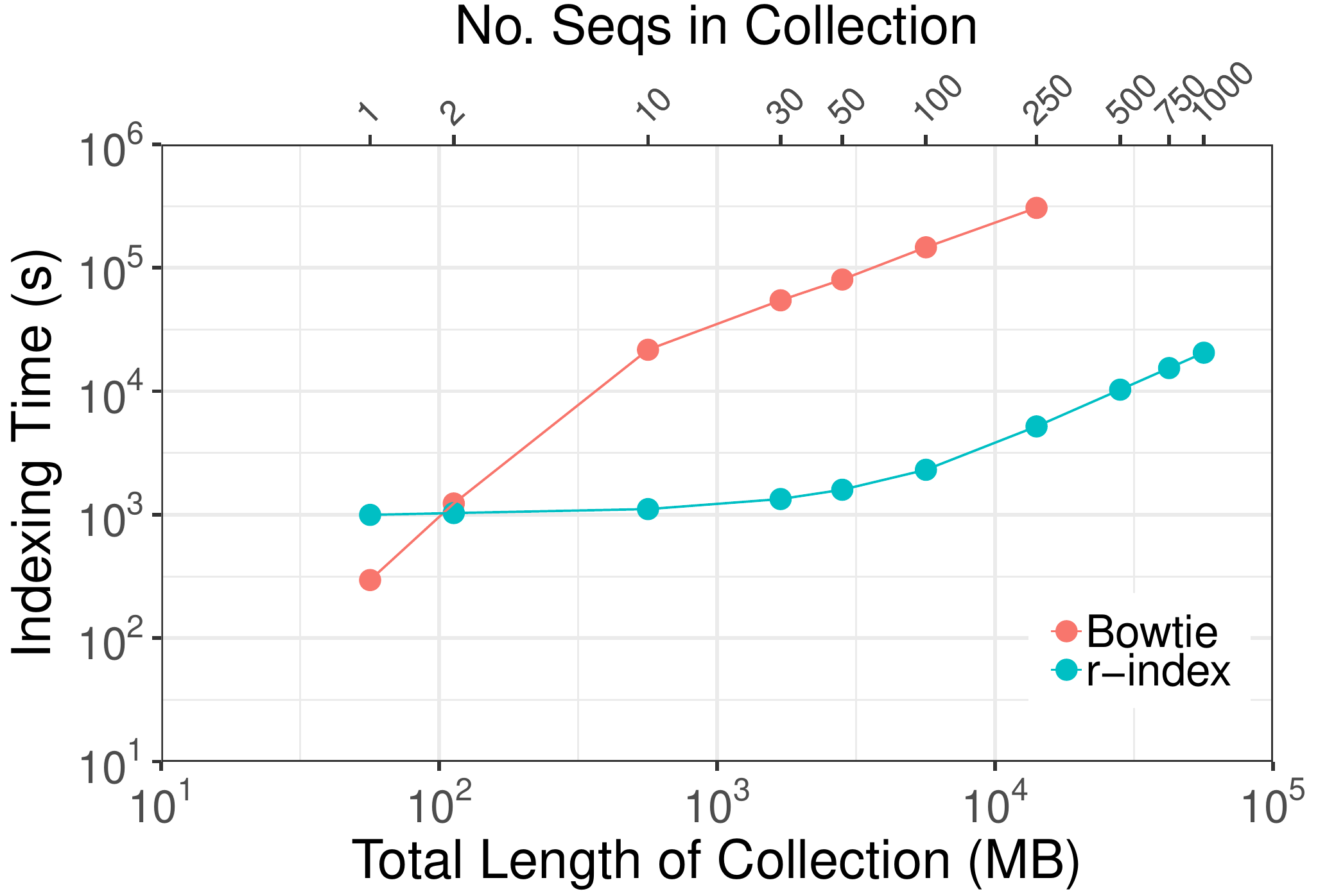}
      \label{fig:constr-time}
  }
  \subfigure[]{
    \includegraphics[width=0.23\textwidth]{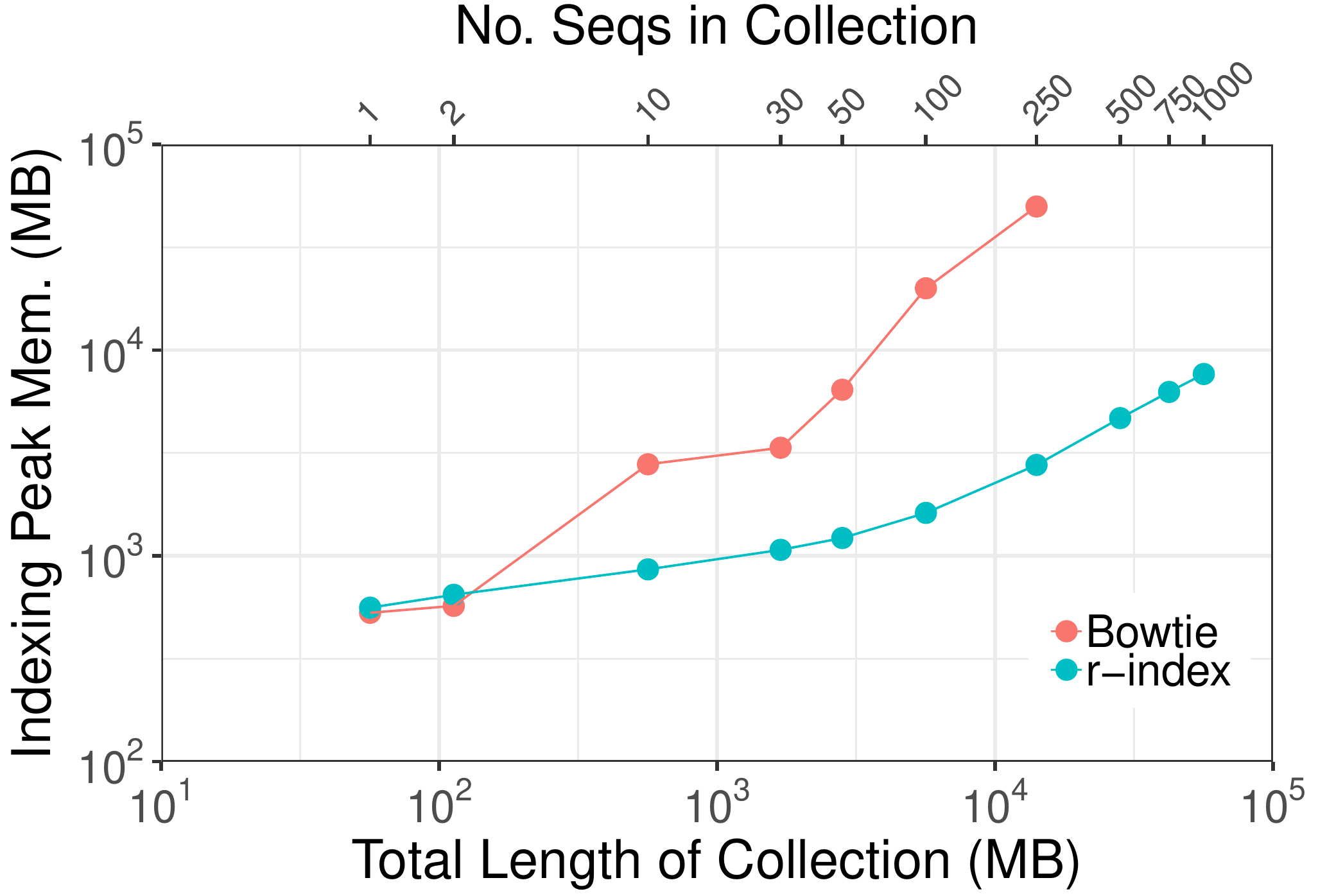}
    \label{fig:constr-mem}
  }
  \subfigure[]{
    \includegraphics[width=0.23\textwidth]{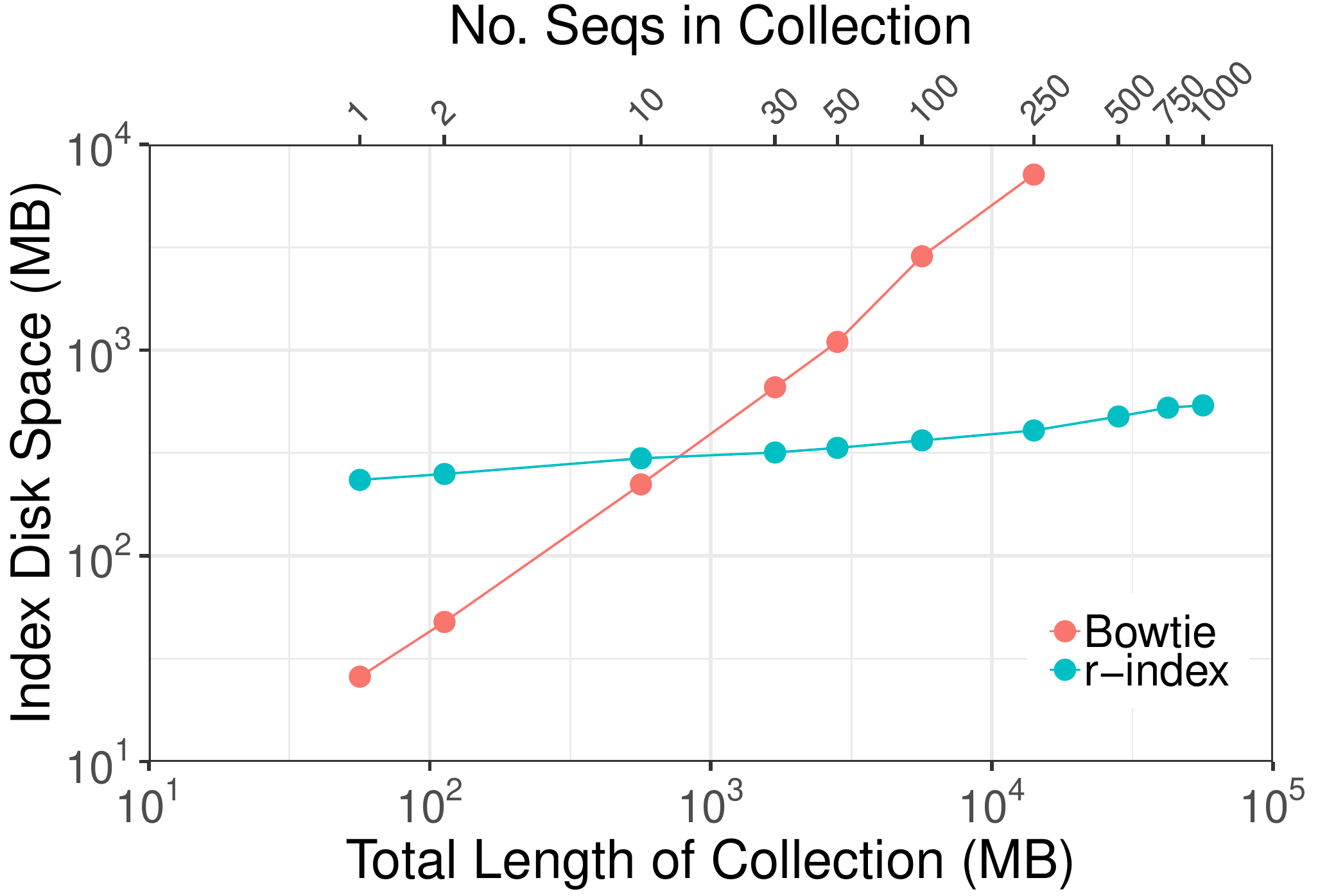}
    \label{fig:index-space}
  }
  \subfigure[]{
    \includegraphics[width=0.23\textwidth]{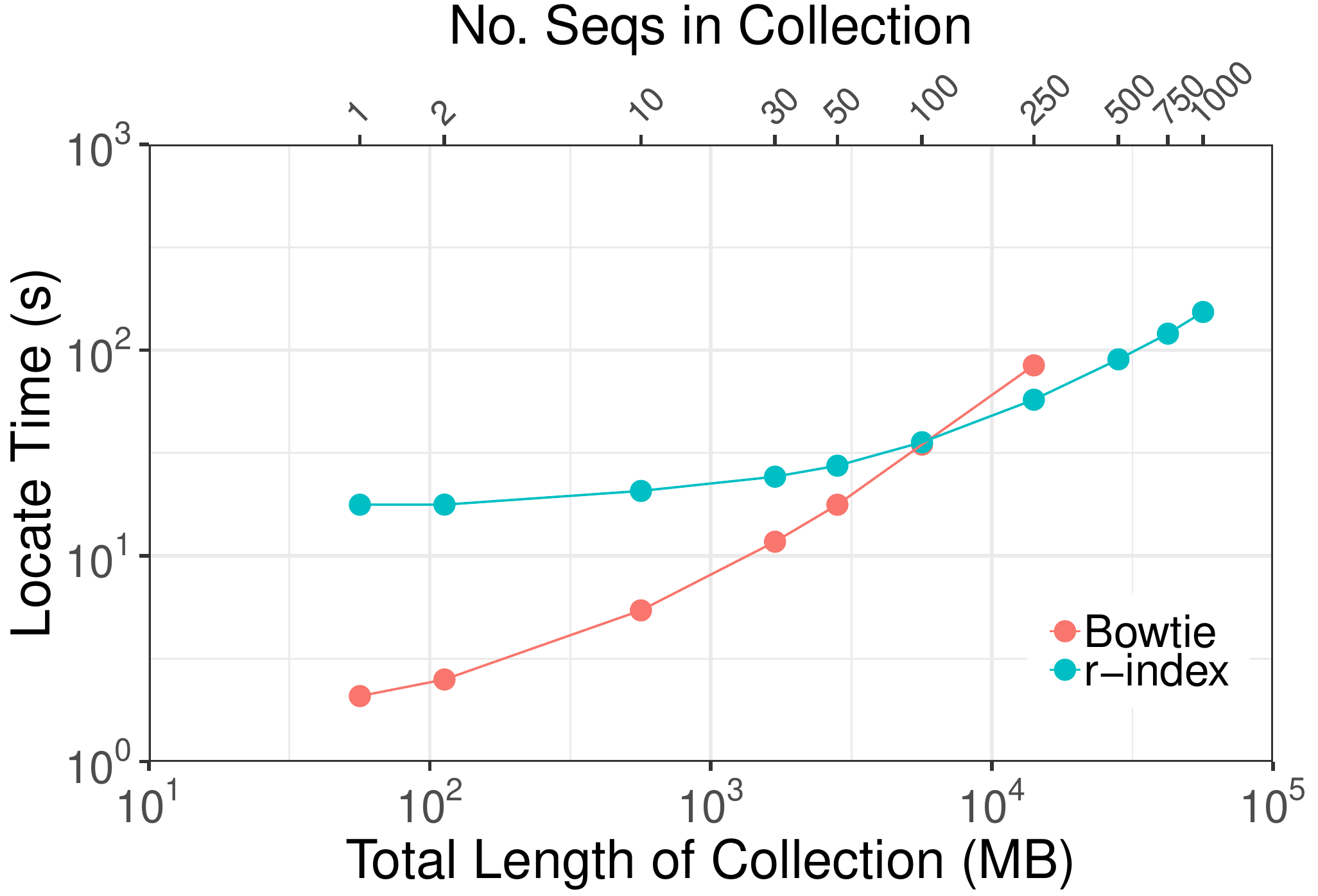}
    \label{fig:locate-time}
  }
  \caption{
      Scalability of r-index and bowtie indexes against chr19 haplotype
      collection size and total sequence length (megabases) with respect to
      index construction time (seconds) (a), index construction peak memory
      (megabytes) (b), index disk space (megabytes) (c), and locate time
      (seconds) of 100,000 100bp queries (d).
      %Bowtie exausts memory and fails
      %after 250 chr19s.  Only \texttt{*.1.ebwt} and \texttt{*.2.ebwt} files are included in the space
      %measurement (c) for Bowtie, for comparability to $r$-index.
      %When locate querying (d), for a collection of \textit{n} sequences, up to \textit{n} locations are found
      %and reported (Bowtie: \texttt{-k = n}, $r$-index: \texttt{--max-hits = n}).
  }
    \label{fig:bowtie-figs}
\end{figure}

First we constructed $r$-index and Bowtie indexes on successively larger chromosome 19
collections (Figure \ref{fig:constr-time}, \ref{fig:constr-mem}). The $r$-index's peak memory is substantially smaller than Bowtie's for larger collections, and
the gap grows with the collection size. At 250 chr19s,
the $r$-index procedure takes about 2\% of the time and 6\%
the peak memory of Bowtie's procedure.
Bowtie fails to construct collections of more than 250
sequences due to memory exhaustion.

Next, we compared the disk footprint of the index files produced by Bowtie and $r$-index
(Figure \ref{fig:index-space}).
The $r$-index currently stores only the forward
strand of the sequence, while the Bowtie index stores both the forward
sequence and its reverse as needed by its double-indexing heuristic \cite{bowtie}.
Since the heuristic is relevant only for approximate matching, we
omit the reverse sequence in these size comparisons.
We also omit the 2-bit encoding of the original text
(in the \texttt{*.3.ebwt} and \texttt{*.4.ebwt} files) as these too are
used only for approximate matching.
Specifically, the Bowtie index size was calculated by adding the sizes of the forward \texttt{*.1.ebwt} and
  \texttt{*.2.ebwt} files, which contain the $\BWT$, $\SA$ sample, and auxiliary data structures for the forward sequence. The size of the $r$-index increased more slowly than Bowtie's, though
the $r$-index was larger for the smallest collections.
This is because, unlike Bowtie which samples a constant fraction
of the SA elements (every 32nd by default), the density of the $r$-index SA sample depends on the
ratio $n/r$.  When the collection is small, $n/r$ is small and more SA
samples must be stored per base.
At 250 sequences, the $r$-index index takes 6\% the space of the Bowtie index.
%complement as well as the two-bit representation of the original index. For fair
%comparison, we leave out the reverse complement and two-bit string from 
%the Bowtie measurements.

% \begin{SCfigure}
%     \centering
%   \subfigure[Locate Time]{
%     \includegraphics[width=0.5\textwidth]{100bp_keqn.pdf}
%   }
%   \caption{Time required to locate 100,000 100-character strings as a function of the
%       number of chr19s indexed (single trials) and total sequence size
%       (megabytes).  For a collection of \textit{n} sequences, up to \textit{n} locations are found
%       and reported (Bowtie: \texttt{-k = n}, $r$-index: \texttt{--max-hits = n}).
%       %Each locate query reports all occurrences of the string, stopping after x
%       %occurrences. 
%       Though Bowtie is faster for smaller
%       collections, the $r$-index becomes faster at 250 chr19s.  Bowtie was unable to
%       build indexes for collections larger than 250 chr19s. 
%   }
%       \label{fig:locate-time}
% \end{SCfigure}

We then compared the speed of the locate query for $r$-index and Bowtie. 
We extracted 100,000 100-character substrings from the chr19 collection
of size 1, which is also contained in all larger collections. We queried these against
both the Bowtie and $r$-indexes.  We used the \texttt{--max-hits} option for $r$-index and the \texttt{-k}
option for Bowtie to set the maximum number of hits reported to be equal to
the collection size.  The actual number of hits reported will often equal this
number, but could be smaller (if the substring differs between individuals
due to genetic variation) or larger (if the substring is from a repetitive portion of the genome).
Since the source of the substrings is present in all the collections, every
query is guaranteed to match at least once. As seen in Figure \ref{fig:locate-time}, the $r$-index locate
query was faster for the collection of 250 chr19s.  No comparison was
possible for larger collections because Bowtie could not build the indexes.

%, indicating to both programs
%that the number of locations to be resolved per query was to be bounded
%by the collection size, and all of those locations up to that bound should be
%reported in the final output. 
%This is because, in a locate query against
%repetitive collections like these haplotypes, one would want to capture the
%query's location in all sequence possible in order to maximize information
%about the query.

\subsection{Indexing whole human genomes} 

%The above experiments measure the efficiency of RLFM on chr19, whereas a more practical
Lastly, we used $r$-index
to index many human genomes at once.
We repeated our measurements for successively larger collections of (concatenated) genomes.
Thus, we first evaluated a series of haplotypes extracted from the 1000 Genomes Project \cite{1kg} phase 3 callset (1KG).
These collections ranged from 1 up to 10 genomes.
%benefit would be to index and query whole genomes. Here, we scale RLFM index
%construction to multiple whole human genome representation using data from the
%1000 Genome Project \cite{1kg} and from long-read \emph{de novo} assemblies found in the
%literature.
%The 1000 Genomes Project provides small variant information for over 2,500
%individuals, ie.  
As the first genome, we selected the GRCh37 reference itself.
For the remaining 9, we used \texttt{bcftools consensus} to insert SNVs
and other variants called by the 1000 Genomes Project for a single haplotype into the GRCh37 reference.
%But the genomes added after that, up to the 10th, were each 
%SNVs and small indels that are mapped to the GRCh37 human
%reference. Like in the above experiments, we generate collections of genomes by
%using the  tool \cite{bcftools} to insert variants
%back in to the reference for each haplotype. We generated 10 genomes in total,
%including the GRCh37 reference, and scale the RLFM according to collection
%size.

Second, we evaluated a series of whole-human genome assemblies from 6 different long-read assembly projects (``LRA'').
We selected GRCh37 reference as the first genome, so that the first data point
would coincide with that of the previous series.
We then added long-read assemblies from
a Chinese genome assembly project \cite{shi_lra},
a Korean genome assembly project \cite{seo_lra}
a project to assemble the well-studied NA12878 individual \cite{jain_lra},
a hydatidiform mole (known as CHM1) assembly project \cite{steinberg_lra}
and the Celera human genome project \cite{huref}.
Compared to the series with only 1000 Genomes Project individuals,
this series allowed us to measure scaling while capturing a wider
range of genetic variation between humans.
This is important since \textit{de novo}
human assembly projects regularly produce assemblies that differ from
the human genome reference by megabases of sequence
(12 megabases in the case of the Chinese assembly \cite{shi_lra}),
likely due to prevalent but hard-to-profile large-scale structural variation.
Such variation was not comprehensively profiled in the 1000 Genomes Project,
which relied on short reads.

The 1KG and LRA series were evaluated twice, once on the forward genome sequences
and once on both the forward and reverse-complement sequences.
This accounts for the fact that different \textit{de novo} assemblies
make different decisions about how to orient contigs.
The $r$-index method achieves compression
only with respect to the forward-oriented versions of the sequences indexed.
That is, if two contigs are reverse complements of each other but otherwise identical, $r$-index
achieves less compression than if their orientations matched.
A more practical approach would be to index both forward and
reverse-complement sequences, as Bowtie 2
\cite{langmead_fast_2012} and BWA \cite{li_aligning_2013} do.

%Incorporating long-read technology into assemblies gives the advantage of
%exposing large scale structural variations within the genome. Some of these
%variations can contain up to 12 megabases of novel sequences that are not
%already present in the reference \cite{shi_lra}, which is harder to capture
%using the small variants provided by 1KG. Also included in this collection is
%the ``Huref'' genome, an assembly generated using Sanger sequencing (which
%can generate reads longer than 500bp) before the development of
%modern long-read technology.

\begin{figure}[t]
\CenterFloatBoxes
\begin{floatrow}
\ttabbox
{\begin{tabular}{crrrrrrr}
\          & \multicolumn{3}{c}{Sequence}    & & \multicolumn{3}{c}{}     \\
\# Genomes & \multicolumn{3}{c}{Length (MB)} & & \multicolumn{3}{c}{$n/r$}  \\ \cline{2-4} \cline{6-8}
           & 1KG        &         & LRA       & & 1KG  &   & LRA        \\ \cline{2-2} \cline{4-4} \cline{6-6} \cline{8-8}
1          & 6,072      &         & 6,072    & & 1.86  &  & 1.86       \\
2          & 12,144     &          & 12,484  & & 3.70  &  & 3.58       \\
3          & 18,217     &          & 17,006  & & 5.38  &  & 4.83       \\
4          & 24,408     &         & 22,739   & & 7.13  &  & 6.25       \\
5          & 30,480     &         & 28,732   & & 8.87  &  & 7.80       \\
6          & 36,671     &         & 34,420   & & 10.63 &  & 9.28
\end{tabular}}
{\caption{Sequence length and $n/r$ statistic with respect to number of whole
genomes for the first 6 collections in the 1000 Genomes (1KG) and long-read assembly (LRA) series.} %These collections included both forward and reverse-complement sequences for each genome.
\label{table:wg-nr}}
\ffigbox
{\includegraphics[width=0.4\textwidth]{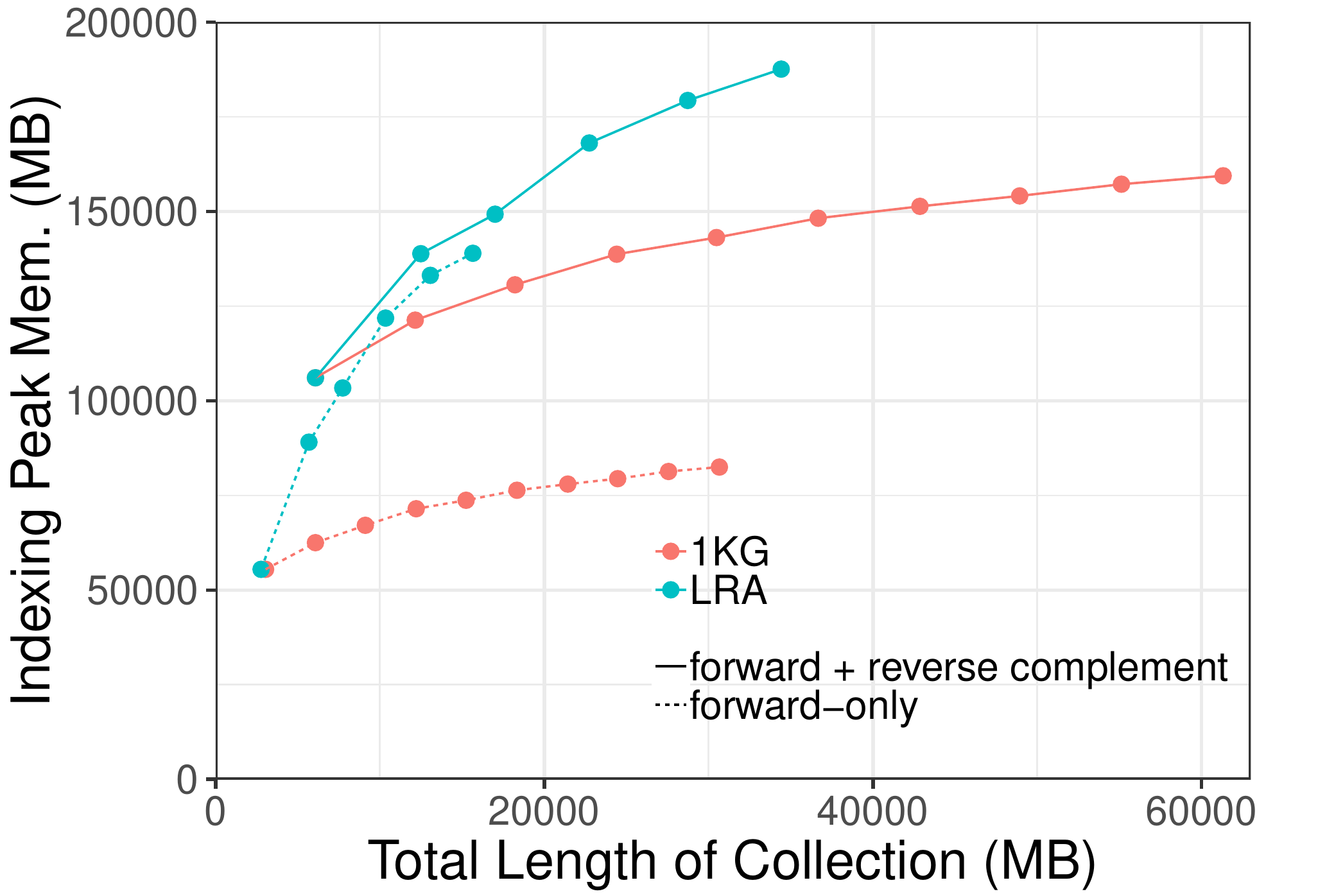}}
{\caption{Peak index-building memory for $r$-index when indexing successively large collections of 1000-Genomes individuals (1KG) and long-read whole-genome assemblies (LRA).}
\label{wg-nr-fig}}
\end{floatrow}
\end{figure}

We measured the peak memory footprint when indexing these collections
(Figure \ref{wg-nr-fig}).
We ran these experiments on an Intel(R) Xeon(R) CPU E5-2650 v4 @ 2.20GHz system with 256GB memory.
% describes the scaling of the time and memory
%footprint of constructing the RLFM index of both the 1000 Genomes and
%long-read-assembly collections with respect to the total length of the
%collection.
Memory footprints for LRA grew more quickly than those for 1KG.
This was expected due to the greater genetic diversity captured in the assemblies.
This may also be due in part to the presence of sequencing errors in the long-read
assembles; long-read technologies are more prone
to indel errors than short-read technologies, for examples, and some may survive in the assemblies.
Also as expected, memory footprints for the LRA series that included both forward
and reverse complement sequences grew more slowly than when just the
forward sequence was included.  This is due to sequences
that differ only (or primarily) in their orientation between assemblies.  All series exhibit
sublinear trends, highlighting the efficacy of $r$-index compression even when indexing
genetically diverse whole-genome assemblies.
Indexing the forward and reverse complement strands of 10 1KG individuals took about 6 hours and 20 minutes
and the final index size was 36GB.  

We also measured lengths and $n/r$ ratios for each collection of whole genomes
(Table \ref{table:wg-nr}).  Consistent with the memory-scaling results, we see that
the $n/r$ ratios are somewhat lower for the LRA series than for the 1KG series,
likely due to greater genetic diversity in the assemblies.

% It must be noted that indexing both strands of 10 1KG individuals
% fails due to memory exhaustion, but for 9 1KG individuals, both strands, indexing takes about 
% $x$ hours and $y$ minutes and the final index size is $z$ gb.

%This
%in turn cause of the larger memory footprint when building the index.

%  higher footprints than the
%corresponding 1KG collections, but both seem to scale sub-linearly with the
%total length. Additionally, the long-read assembly collections scale
%less efficiently in terms of $n/r$ than do the 1KG collections.

%The discrepancy between the scaling and construction footprints 
%of the two collections can be explained by both the the novel sequence included
%in the assemblies and the variable qualities of these assemblies.  Some
%assemblies are not resolved at the chromosome level and exist as scaffolds and
%contigs, and others may not have went through rigorous error correction
%procedures.  Additionally, these long-read assemblies are not guaranteed to be
%of the same strand as the reference; they might instead contain the reverse
%complement. These caveats, in addition  to any structural variations that the
%assemblies might capture, would slow the increase of the $n/r$ statistic with
%the addition of sequences to the collection. 

%%% Local Variables:
%%% mode: latex
%%% TeX-master: "recomb"
%%% End:

\section{Conclusions and Future Work} \label{sec:future}
We give an algorithm for building the $\SA$ and $\SA$ sample from the prefix-free parse of an input string $S$, which fully completes the practical challenge of building the index proposed by Gagie et al. \cite{Travis18}. 
This leads to a mechanism for building a complete index of large databases --- which is the linchpin in developing practical means for pan-genomics short read alignment.  In fact, we apply our method for indexing partial and whole human genomes, and show that it scales better than Bowtie with respect to both memory and time. This allows for an index to be constructed for large collections of chromosome 19s (500 or more); a task that is out of reach of Bowtie --- as  exceeded our limit of 512 GB of memory. 

Even though this work opens up doors to indexing large collections of genomes, it also highlights problems that warrant further investigation.  For example, there still remains a significant amount of work in adapting the index to work well on large sets of sequence reads.  This problem not only requires the construction of the $r$-index but also an efficient means to update the index as new datasets become available.  Moreover, there is interest in supporting more sophisticated queries than just pattern matching, which would allow for more complex searches of large databases.  

%it remains open as to how to compute a mixed $\SA$ sample that uses regular sampling sometimes and the $r$-index sometimes, depending on distance between elements in the predecessor structure.  From a more applied perspective

%%% Local Variables:
%%% mode: latex
%%% TeX-master: "recomb"
%%% End:


\newpage

\begin{thebibliography}{10}

\bibitem{BGI18}
H.~Bannai, T.~Gagie, and T.~I.
\newblock Online {LZ77} parsing and matching statistics with {RLBWTs}.
\newblock In {\em Proceedings of tjhe 29th Annual Symposium on Combinatorial
  Pattern Matching (CPM)}, volume 105, pages 7:1--7:12, 2018.

\bibitem{boucher2018}
C.~Boucher, T.~Gagie, A.~Kuhnle, and G.~Manzini.
\newblock Prefix-free parsing for building big {BWTs}.
\newblock In {\em Proceedings of 18th International Workshop on Algorithms in
  Bioinformatics (WABI)}, volume 113, pages 2:1--2:16, 2018.

\bibitem{Burrows18}
M.~Burrows and D.J. Wheeler.
\newblock A block sorting lossless data compression algorithm.
\newblock Technical Report 124, Digital Equipment Corporation, 1994.

\bibitem{1kg}
The 1000 Genomes~Project Consortium.
\newblock A global reference for human genetic variation.
\newblock {\em Nature}, 526(7571):68--74, October 2015.

\bibitem{Danek2014}
A.~Danek, S.~Deorowicz, and S.~Grabowski.
\newblock {Indexes of large genome collections on a PC}.
\newblock {\em PLoS ONE}, 9(10), 2014.

\bibitem{Deorowicz2015}
S.~Deorowicz, M.~Kokot, S.~Grabowski, and A.~Debudaj-Grabysz.
\newblock {KMC 2: Fast and resource-frugal k-mer counting}.
\newblock {\em Bioinformatics}, 31(10):1569--1576, 2015.

\bibitem{Garrison}
Garrison E. et~al.
\newblock Variation graph toolkit improves read mapping by representing genetic
  variation in the reference.
\newblock {\em Nature Biotechnology}, 36(9):875--879, 2018.

\bibitem{Ferrada2014}
H.~Ferrada, T.~Gagie, T.~Hirvola, and S.~J. Puglisi.
\newblock {Hybrid indexes for repetitive datasets}.
\newblock {\em Philosophical Transactions of the Royal Society A: Mathematical,
  Physical and Engineering Sciences}, 372(2016):1--9, 2014.

\bibitem{Ferrada2018}
H.~Ferrada, D.~Kempa, and S.J. Puglisi.
\newblock {Hybrid Indexing Revisited}.
\newblock In {\em Proceedings of the 21st Algorithm Engineering and Experiments
  (ALENEX)}, pages 1--8, 2018.

\bibitem{FM05}
P.~Ferragina and G.~Manzini.
\newblock Opportunistic data structures with applications.
\newblock In {\em Proceedings of the 41st Annual Symposium on Foundations of
  Computer Science (FOCS)}, pages 390--398, 2000.

\bibitem{Travis18}
T.~Gagie, G.~Navarro, and N.~Prezza.
\newblock Optimal-time text indexing in bwt-runs bounded space.
\newblock In {\em Proceedings of the 29th Annual Symposium on Discrete
  Algorithms (SODA)}, pages 1459--1477, 2018.

\bibitem{Gagie2015}
T.~Gagie and S.J. Puglisi.
\newblock {Searching and Indexing Genomic Databases via Kernelization}.
\newblock {\em Frontiers in Bioengineering and Biotechnology}, 3:10--13, 2015.

\bibitem{bwa}
Li~H. and Durbin R.
\newblock {Fast and accurate short read alignment with Burrows-Wheeler
  Transform}.
\newblock {\em Bioinformatics}, 25:1754--60, 2009.

\bibitem{Huang2013a}
L.~Huang, V.~Popic, and S.~Batzoglou.
\newblock {Short read alignment with populations of genomes}.
\newblock 29(13), 2013.

\bibitem{jain_lra}
M.~Jain et~al.
\newblock Nanopore sequencing and assembly of a human genome with ultra-long
  reads.
\newblock {\em Nature Biotechnology}, 36(4):338--345, April 2018.

\bibitem{seo_lra}
S.~Jeong-Sun et~al.
\newblock \textit{{De} novo} assembly and phasing of a {Korean} human genome.
\newblock {\em Nature}, 538(7624):243--247, October 2016.

\bibitem{Karkkainen2015}
J.~K{\"{a}}rkk{\"{a}}inen, D.~Kempa, and S.~J. Puglisi.
\newblock Parallel external memory suffix sorting.
\newblock In {\em Proceedings of the 26th Annual Symposium on Combinatorial
  Pattern Matching (CPM)}, pages 329--342, 2015.

\bibitem{langmead_fast_2012}
B.~Langmead and S.L. Salzberg.
\newblock Fast gapped-read alignment with {Bowtie} 2.
\newblock {\em Nature Methods}, 9(4):357, March 2012.

\bibitem{bowtie}
B.~Langmead, C.~Trapnell, M.~Pop, and S.~L. Salzberg.
\newblock Ultrafast and memory-efficient alignment of short {DNA} sequences to
  the human genome.
\newblock {\em Genome Biology}, 10, 2008.

\bibitem{huref}
S.~Levy et~al.
\newblock The {Diploid} {Genome} {Sequence} of an {Individual} {Human}.
\newblock {\em PLoS Biology}, 5(10):e254, September 2007.

\bibitem{soap}
C.~Li, R.and~Yu, Y.~Li, T.-W. Lam, S.-M. Yiu, K.~Kristiansen, and J.~Wang.
\newblock Soap2: an improved tool for short read alignment.
\newblock {\em Bioinformatics}, 25(15):1966--1967, 2009.

\bibitem{li_aligning_2013}
Heng Li.
\newblock Aligning sequence reads, clone sequences and assembly contigs with
  {BWA}-{MEM}.
\newblock {\em arXiv:1303.3997 [q-bio]}, March 2013.
\newblock arXiv: 1303.3997.

\bibitem{Maciuca2016}
S.~Maciuca, C.~{del Ojo Elias}, G.~McVean, and Z.~Iqbal.
\newblock A natural encoding of genetic variation in a {B}urrows-{W}heeler
  transform to enable mapping and genome inference.
\newblock In {\em Proceedings of the 16th Annual Workshop on Algorithms in
  Bioinformatics (WABI)}, pages 222--233, 2016.

\bibitem{Makinen2010}
V.~M{\"{a}}kinen, G.~Navarro, J.~Sir{\'{e}}n, and N.~V{\"{a}}lim{\"{a}}ki.
\newblock {Storage and retrieval of highly repetitive sequence collections.}
\newblock {\em Journal of Computational Biology}, 17(3):281--308, 2010.

\bibitem{Peng2012}
Yu~Peng, Henry C~M Leung, S.~M. Yiu, and Francis Y~L Chin.
\newblock {IDBA-UD: A de novo assembler for single-cell and metagenomic
  sequencing data with highly uneven depth}.
\newblock {\em Bioinformatics}, 28(11):1420--1428, 2012.

\bibitem{PP18}
A.~Policriti and N.~Prezza.
\newblock {LZ77} computation based on the run-length encoded {BWT}.
\newblock {\em Algorithmica}, 80(7):1986--2011, 2018.

\bibitem{Schneeberger2009}
K.~Schneeberger et~al.
\newblock {Simultaneous alignment of short reads against multiple genomes}.
\newblock {\em Genome Biology}, 10(9), 2009.

\bibitem{shi_lra}
L.~Shi et~al.
\newblock Long-read sequencing and \textit{de novo} assembly of a {Chinese}
  genome.
\newblock {\em Nature Communications}, 7:12065, June 2016.

\bibitem{Siren2014}
J.~Sir{\'e}n, N.~V{\"a}lim{\"a}ki, and V.~M{\"a}kinen.
\newblock Indexing graphs for path queries with applications in genome
  research.
\newblock 11(2):375--388, 2014.

\bibitem{steinberg_lra}
K.M. Steinberg et~al.
\newblock Single haplotype assembly of the human genome from a hydatidiform
  mole.
\newblock {\em Genome Research}, page gr.180893.114, November 2014.

\bibitem{STBASBM17}
E.L. Stevens, R.~Timme, E.W. Brown, M.W. Allard, E.~Strain, K.~Bunning, and
  S.~Musser.
\newblock The public health impact of a publically available, environmental
  database of microbial genomes.
\newblock {\em Frontiers in Microbiology}, 8:808, 2017.

\bibitem{Valenzuela2017}
D.~Valenzuela and V.~M{\"a}kinen.
\newblock {CHIC: a short read aligner for pan-genomic references}.
\newblock {\em BMC Bioinformatics}, 19(Suppl 2):87, 2018.

\bibitem{Valenzuela2018}
D.~Valenzuela, T.~Norri, N.~V{\"a}lim{\"a}ki, E.~Pitk{\"a}nen, and
  V.~M{\"a}kinen.
\newblock Towards pan-genome read alignment to improve variation calling.
\newblock {\em BMC Genomics}, 19(2):87, 2018.

\bibitem{Wandelt2013}
S.~Wandelt, J.~Starlinger, M.~Bux, and U.~Leser.
\newblock {RCSI: Scalable similarity search in thousand(s) of genomes}.
\newblock {\em Proceedings of the VLDB Endowment}, 6(13):1534--1545, 2013.

\end{thebibliography}
\end{document}